\documentclass[useAMS,usenatbib,usegraphicx]{mn2e}
\usepackage{longtable}
\usepackage{amsmath}
\usepackage{amssymb}
\usepackage{longtable}
\usepackage{latexsym}

\usepackage{lscape}
\usepackage{longtable}
\usepackage{rotating}           % for sideways tables/figures
\usepackage{hyperref}

\usepackage{graphics}

\usepackage{color}

\usepackage[normalem]{ulem}

\def\hatlas{{\it H}-ATLAS}

\def\LIR{$L_{\rm IR}$}
\def\Cii{[C\,{\sc ii}]}

\def\gs{\mathrel{\raise0.35ex\hbox{$\scriptstyle >$}\kern-0.6em
\lower0.40ex\hbox{{$\scriptstyle \sim$}}}}
\def\ls{\mathrel{\raise0.35ex\hbox{$\scriptstyle <$}\kern-0.6em
\lower0.40ex\hbox{{$\scriptstyle \sim$}}}}
\def\m@th{\mathsurround=0pt }
\def\eqalign#1{\null\,\vcenter{\openup1\jot \m@th
 \ialign{\strut\hfil$\displaystyle{##}$&$\displaystyle{{}##}$\hfil
 \crcr#1\crcr}}\,}

\long\def\symbolfootnote[#1]#2{\begingroup%
  \def\thefootnote{\fnsymbol{footnote}}\footnote[#1]{#2}\endgroup}

\title[A multi-wavelength exploration of the \Cii/IR ratio in
  {\it H}-ATLAS/GAMA galaxies out to $z=0.2$]
      {A multi-wavelength exploration of the \Cii/IR ratio in
        \textit{\textbf{H}}-ATLAS/GAMA galaxies out to $\mathbf{z=0.2}$}
\author[E.~Ibar et al.]
{\parbox{\textwidth}{\raggedright E.~Ibar,$^{1}$\thanks{E-mail: \texttt{eduardo.ibar@uv.cl}}
M.A.~Lara-L{\'o}pez,$^{2,3}$
R.~Herrera-Camus,$^{4}$
R.~Hopwood,$^{5}$
A.~Bauer,$^{3}$
R.J.~Ivison,$^{6,7}$
M.J.~Micha{\l}owski,$^{7}$
H.~Dannerbauer,$^{8}$
P.~van der Werf,$^{9}$
D.~Riechers,$^{10,11}$
N.~Bourne,$^{7}$
M.~Baes,$^{12}$
I.~Valtchanov,$^{13}$
L.~Dunne,$^{14,7}$
A.~Verma,$^{15}$
S.~Brough,$^{3}$
A.~Cooray,$^{16,10}$
G.~De Zotti,$^{17,18}$
S.~Dye,$^{19}$
S.~Eales,$^{20}$
C.~Furlanetto,$^{19,21}$
S.~Maddox,$^{14,7}$
M.~Smith,$^{20}$
O.~Steele,$^{22}$
D.~Thomas$^{22}$ and
E.~Valiante$^{20}$}\vspace{0.4cm}\\
\parbox{\textwidth}{\raggedright $^{1}$Instituto de F\'isica y Astronom\'ia, Universidad de Valpara\'iso, Avda.\ Gran Breta\~na 1111, Valparaiso, Chile\\
$^{2}$Instituto de Astronom\'ia, Universidad Nacional Aut\'onoma de M\'exico, A.P. 70-264, 04510 M\'exico, D.F., M\'exico\\
$^{3}$Australian Astronomical Observatory, PO Box 915, North Ryde, NSW 1670, Australia\\
$^{4}$Department of Astronomy, University of Maryland, College Park, MD 20742-2421, USA\\
$^{5}$Astrophysics Group, Imperial College London, Blackett Laboratory, Prince Consort Road, London SW7 2AZ, UK\\
$^{6}$European Southern Observatory, Karl-Schwarzschild-Stra{\ss}e 2, 85748 Garching bei M\"unchen, Germany\\
$^{7}$Institute for Astronomy, University of Edinburgh, Royal Observatory, Blackford Hill, Edinburgh EH9 3HJ, UK\\
$^{8}$Universit\"at Wien, Institut f\"ur Astrophysik, T\"urkenschanzstra\ss e 17, 1180 Vienna, Austria\\
$^{9}$Leiden Observatory, Leiden University, P.O.\ Box 9513, NL-2300 RA Leiden, The Netherlands\\
$^{10}$California Institute of Technology, 1200 E. California Blvd., Pasadena, CA 91125, USA\\
$^{11}$Department of Astronomy, Cornell University, 220 Space Sciences Building, Ithaca, NY 14853, USA\\
$^{12}$Sterrenkundig Observatorium, Universiteit Gent, Krijgslaan 281 S9, B-9000 Gent, Belgium\\
$^{13}$Herschel Science Centre, European Space Astronomy Centre, Villanueva de la Ca\~nada, 28691 Madrid, Spain\\
$^{14}$Department of Physics and Astronomy, University of Canterbury, Private Bag 4800, Christchurch 8140, New Zealand\\
$^{15}$Department of Astrophysics, Denys Wilkinson Building, University of Oxford, Keble Road, Oxford OX1 3RH, UK\\
$^{16}$Dept. of Physics \& Astronomy, University of California, Irvine, CA 92697, USA\\
$^{17}$INAF - Osservatorio Astronomico di Padova, Vicolo dell'Osservatorio 5, I-35122 Padova, Italy.\\
$^{18}$SISSA, Via Bonomea 265, I-34136 Trieste, Italy\\
$^{19}$School of Physics and Astronomy, University of Nottingham, NG7 2RD, UK\\
$^{20}$School of Physics and Astronomy, Cardiff University, Queens Buildings, The Parade, Cardiff CF24 3AA, UK\\
$^{21}$CAPES Foundation, Ministry of Education of Brazil, Brasilia/DF, 70040-020, Brazil\\
$^{22}$Institute of Cosmology and Gravitation, University of Portsmouth, Dennis Sciama Building, Burnaby Road, Portsmouth PO1 3FX, UK}}
\begin{document}

\date{Accepted 24 February 2015. Received 17 February 2015; in original form 17 October 2014}
%\date{Accepted --. Received --; in original form --}

\pagerange{\pageref{firstpage}--\pageref{lastpage}} \pubyear{2015}

\maketitle

\label{firstpage}

\begin{abstract}
  We explore the behaviour of
    [C\,{\sc{ii}}]\,$\lambda$157.74\,$\mu$m forbidden fine-structure
    line observed in a sample of 28 galaxies selected from
    $\sim$\,50\,deg$^2$ of the {\it H}-ATLAS survey. The sample is
    restricted to galaxies with flux densities higher than $S_{\rm
      160\mu m}>150\,$mJy and optical spectra from the GAMA survey at
    $0.02<z<0.2$. Far-IR spectra centred on this redshifted line were
    taken with the PACS instrument on-board the {\it Herschel Space
      Observatory}. The galaxies span $10<{\rm log} (L_{\rm IR}/{\rm
      L}_\odot) <12$ (where $L_{\rm IR}\equiv L_{\rm IR}[8-1000\mu{\rm
        m}]$) and $7.3<{\rm log} (L_{\rm [CII]}/{\rm L}_\odot) <9.3$,
    covering a variety of optical galaxy morphologies. The sample
    exhibits the so-called \Cii\ deficit at high IR luminosities,
    i.e.\ $L_{[{\rm C\,{\sc II}}]}$/\LIR\ (hereafter \Cii/IR)
    decreases at high \LIR. We find significant differences between
    those galaxies presenting \Cii/IR\,$>2.5\,\times\,10^{-3}$ with
    respect to those showing lower ratios. In particular, those with
    high ratios tend to have: (1) \LIR\,$< 10^{11}$\,L$_\odot$; (2)
    cold dust temperatures, $T_{\rm d}<30$\,K; (3) disk-like
    morphologies in $r$-band images; (4) a {\it WISE} colour $0.5\ls
    S_{\rm 12\mu m}/S_{\rm 22\mu m}\ls1.0$; (5) low surface brightness
    $\Sigma_{\rm IR}\approx10^{8-9}\,L_\odot\,$kpc$^{-2}$, (6) and
    specific star-formation rates of
    sSFR\,$\approx0.05-3$\,Gyr$^{-1}$. We suggest that the strength of
    the far-UV radiation fields ($<$$G_{\rm O}$$>$) is main parameter
    responsible for controlling the \Cii/IR ratio. It is possible that
    relatively high $<$$G_{\rm O}$$>$ creates a positively charged
    dust grain distribution, impeding an efficient photo-electric
    extraction of electrons from these grains to then collisionally
    excite carbon atoms. Within the brighter IR population, $11<{\rm
      log} (L_{\rm IR}/{\rm L}_\odot) <12$, the low \Cii/IR ratio is
    unlikely to be modified by \Cii\ self absorption or controlled by
    the presence of a moderately luminous AGN (identified via the BPT
    diagram).
%The neutral gas is expected to not to dominate
%  the \Cii\ emission in more powerful ULIRG-like galaxies.
\end{abstract}

\begin{keywords}
Galaxies: starburst; ISM lines and bands; ISM
  evolution; Infrared: ISM resolved and unresolved sources as a
  function of wavelength;
\end{keywords}

\section{Introduction}
\label{sec_intro}

Understanding the chemical and physical evolution of a galaxy is far
from trivial. Newly born stars consume and process the available gas,
whilst heating the interstellar medium (ISM), and supernovae enrich
the environment with heavy elements, contributing to potentially
complex feedback processes. A good description of ISM physics under
the influence of stellar radiation fields was achieved using
photo-dissociation region (PDR) modelling \citep[e.g.][]{Tielens85}.
These models can explain the origin of most of the dense ISM emission
from star-forming galaxies, including the major cooling fine-structure
lines of carbon/nitrogen/oxygen (C/N/O) and the underlying
Infrared (IR) continuum emission produced by interstellar dust.

At IR wavelengths, the most prominent emission line is
\Cii\,$\lambda$157.74\,$\mu$m
($^2P_{3/2}\rightarrow$$^2P_{1/2}$; $E_{\rm ul}/k=92\,$K),
which carries $\sim 0.1-1$ per cent of the bolometric power emitted by
star-forming galaxies \citep{Stacey91,Stacey10}, with a \Cii\ flux
typically $1000\times$ that of CO\,($J=1$--0) at 115\,GHz. The low
ionisation potential, 11.26\,eV, makes \Cii\ a key participant in
cooling the warm and diffuse media, converting it into cold and dense
clouds that can then collapse to form stars \citep{Dalgarno72}. As a
fine-structure line, \Cii\ is an excellent tracer of all the different
stages of evolution of the ISM: it can be excited by collisions with
electrons in the warm ionised medium; H\,{\sc i} in the warm or cold
diffuse media; and H$_2$ in the warm and dense molecular gas. Its
intensity is sensitive to the column density, the volume density and
the kinetic temperature of the ISM \citep{Pineda13}.

In the plane of the Milky Way, \citet{Pineda13} show that the
\Cii\ line emission emerges predominantly at Galactocentric distances
between 4 and 10\,kpc. Considering a scale height for the Galaxy,
\citet{Pineda14} finds that the ISM components that contribute to the
\Cii\ luminosity of the Galaxy have roughly comparable contributions:
30 per cent comes from dense PDRs, $\sim$\,20 per cent from ionised
gas, $\sim$\,25 per cent from diffuse atomic gas and $\sim$\,25 per
cent from CO-dark H$_2$ (at the surface of molecular clouds where
carbon is not in the form of CO). In the local spiral galaxy M\,33,
\citet{Kramer13} suggest that the \Cii\ emission related to neutral
gas corresponds to $\sim$\,15\,\% of the total within 2\,kpc from the
galaxy centre, while this percentage seems to increment up to
$\sim$\,40\,$\pm$\,20\,\% in the outer part of the disk (between 2 and
7\,kpc).

In distant galaxies, strong observational limitations impede a
detailed characterisation of the different phases of the ISM.
Typically detected in a single telescope beam, the ISM phases are all
mixed together, hence \Cii\ line detections relate to averaged
quantities of an ensemble of individual PDRs, ionised regions, etc.
\citet{Madden93} suggest that $\sim$\,75\,\% of the \Cii\ emission
from the spiral galaxy NGC\,6946 originates from cold neutral hydrogen
clouds, and $\lesssim\,40\,\%$ from the diffuse galaxy disk
(\citealt{Contursi02}). On the other hand, averaged ISM properties of
luminous- and ultra-luminous- IR galaxies (LIRGs and ULIRGs) are found
to vary considerably with respect to those observed in the Milky
Way. LIRGs/ULIRGs present much higher star-formation rates and
evidence much larger amounts of ionised gas -- up to 50\,\% of the
total (e.g.\ \citealt{Malhotra01}). The rest of their \Cii\ luminosity
is expected to come from dense PDR-related ISM
(e.g.\ \citealt{Negishi01}). At the other extreme, \citet{Cormier12}
estimate that the low metallicity dwarf star-forming galaxy Haro 11
produces only 10\,\% of its \Cii\ emission in PDRs, probably because
radiation fields penetrate deep into the ISM components.

The examples shown above clearly evidence the intricate decomposition
of the \Cii\ emission into the different ISM phases. Great effort has
been dedicated to find \Cii\ correlations with various galaxy
properties. Previous studies have shown that the integrated \Cii\ line
strength depends on the Polycyclic Aromatic Hydrocarbon (PAH) emission
lines (e.g.\ \citealt{Bakes94}), IR colour (\citealt{Malhotra01}), the
degree of active galactic nuclei (AGN; e.g.\ \citealt{Stacey10})
activity, among other various correlations. Those with powerful AGN
show fainter \Cii\ than pure star-forming galaxies, at fixed
\LIR\ (\citealt{Negishi01, Sargsyan12}). \Cii\ has also been used for
diagnostic purposes in galaxies at cosmological distances
\citep[e.g.][]{Ivison10, Valtchanov11, George13} and has sometimes
betrayed a galaxy's redshift \citep[e.g.][]{Swinbank12,
  George13}. Indeed, \Cii\ is a potentially unrivalled
  tracer of the total gas mass or the star-formation rate or the
  dynamics in the most distant galaxies \citep{Maiolino05, deLooze14,
    DeBreuck14, Herrera-Camus15}.

Early observations of local star-forming galaxies with the Kuiper
Airborne Observatory \citep{Stacey91} and the {\it Infrared Space
  Observatory} \citep[ISO;][]{Malhotra97} showed that \Cii/IR is
roughly constant at 1--3\,$\times10^{-3}$ (with a factor of three of
scatter) for \LIR\,$<10^{11}$\,L$_\odot$ galaxies \citep[it is higher
  in low-metallicity environments --][]{Madden00,Rubin09}, although it
drops rapidly at higher \LIR. This behaviour is usually referred to as
the `\Cii\ deficit'. \Cii/IR also correlates strongly with
60\,$\mu$m/100\,$\mu$m colour \citep[a proxy for dust temperature,
  e.g.][]{Diaz-Santos13} and the IR/$B$-band luminosity ratio,
\LIR/$L_{\rm B}$ (\citealt{Malhotra01}). Various explanations for the
\Cii\ deficit have been offered: (1) in high radiation fields,
photoionisation of dust grains might saturate, hence the energy of
their ejected photo-electrons is reduced, reducing also the gas
heating \citep{Luhman03}; (2) the deficit is due to an increase in the
collisional de-excitation of the \Cii\ transition due to an increase
in gas density \citep{Negishi01}; (3) a significant fraction of the IR
emission may arise from dust absorption of photons
  from old stellar populations which are not related directly to PDRs
  \citep[e.g.][]{Rowan-Robinson10}; (4) a significant portion of the
  IR emission could emanate from `dust-bounded' structures within
  photoionised gas regions, but the \Cii\ line does not
  (\citealt{Luhman03,Abel09}); (5) the \Cii\ line may be
self-absorbed or optically thick \citep[e.g.\ NGC\,6334
  --][]{Boreiko95} in highly embedded regions \citep{Fischer10}
(see also \citealt{Gerin15}), or (6) the level of IR
emission in the most IR-luminous objects may be boosted by AGN
activity \citep{Curran09,Sargsyan12}.

Recent evidence, based on observations of local star-forming galaxies
with the {\it Herschel Space Observatory}\footnote{{\it Herschel} is
  an ESA space observatory with science instruments provided by
  European-led Principal Investigator consortia and with important
  participation from NASA.} \citep{Pilbratt10}, points to different
modes of star formation driven by the star-formation efficiency,
\LIR/M$_{\rm H_2}$ \citep[e.g.][]{Young86}, regardless of the origin
of the ionised or neutral phase of the ISM \citep{Gracia-Carpio11}.
\citeauthor{Gracia-Carpio11} find that \LIR/M$_{\rm H_2} \gtrsim
100$\,L$_\odot$\,M$_\odot^{-1}$ marks a point at which the average
properties of the neutral and ionised gas are significantly different.
Using a sample of powerful star-forming galaxies they propose a
scenario in which highly compressed, more efficient star formation,
creates largely enhanced ionisation parameters that manifest
themselves in lower line to continuum ratios. This value of
\LIR/M$_{\rm H_2}$ is closely related to that at
which \citet{Genzel10} and \citet{Daddi10} claim a transition to a
more efficient star-formation mode, above the so-called `main
sequence' for star-forming galaxies \citep{Elbaz11}.

{\it Herschel} was able to observe \Cii\ line emission from many local
galaxies -- a legacy that will last into the era of SPICA \citep[the
  Space Infrared Telescope for Cosmology and Astrophysics
  --][]{Nakagawa12}. Comprehensive analyses of these observations are
mandatory if we are to interpret ground-based observations of
\Cii\ towards high-redshift galaxies, e.g.\ with the Atacama Large
Millimeter Array (ALMA), where the [C\,{\sc ii}] line is shifted into
accessible atmospheric windows \citep[e.g.][]{Maiolino05, Maiolino09,
  Walter09, DeBreuck11, DeBreuck14, Riechers13, Riechers14}. Indeed,
recent studies of luminous, high-redshift galaxies have revealed a
\Cii\ behaviour which contrasts with that seen locally. Instead of
looking like powerful ULIRGs -- at the same luminosity -- they exhibit
striking similarities in terms of \Cii/IR ratios to normal local
star-forming galaxies \citep{Ivison10, Hailey-Dunsheath10, Stacey10,
  Valtchanov11}.

In this paper, we present {\it Herschel} Photodetector
  Array Camera and Spectrometer (PACS; \citealt{Poglitsch10})
Integral Field Unit (IFU) detections of a sample of dusty galaxies.
We populate the $L_{\rm [CII]}$ versus \LIR\ diagram with galaxies
that lie at distances between 90\,Mpc ($z\sim0.02$) and 1000\,Mpc
($z\sim0.2$). This study therefore bridges the gap between local
galaxies and the growing body of data acquired for galaxies at high
redshift. Our study performs a uniquely wide and detailed parameter
space exploration that perfectly complements recent {\it Herschel}
studies at low and intermediate redshifts, e.g.\ the {\it Herschel}
ULIRG Survey (HERUS; \citealt{Farrah13}) and the
Great Observatories All-sky LIRG Survey \citep[GOALS
  --][]{Diaz-Santos13}, a sample of LIRGs at $z<0.09$.

In what follows, we adopt a Kroupa initial mass function \citep[IMF
--][]{Kroupa03} and a $\Lambda$CDM cosmology with
$H_0=70$\,km\,s$^{-1}$\,Mpc$^{-1}$, $\Omega_{\rm M}=0.27$ and
$\Omega_{\Lambda}=0.73$.

\begin{figure}
   \centering
   \includegraphics[scale=0.50]{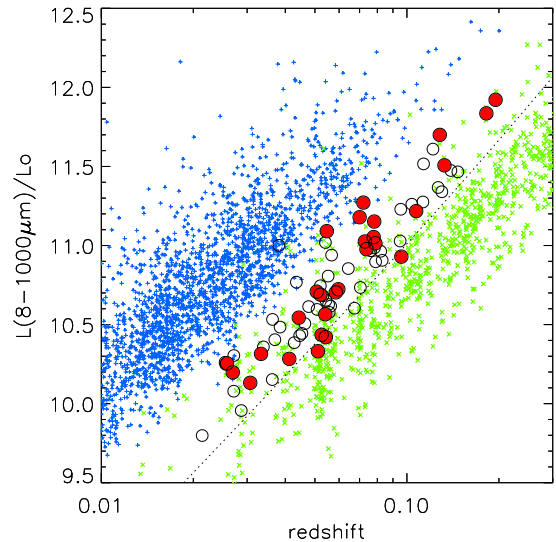}
   \caption{IR luminosity versus redshift for the 28 galaxies that
     were observed with PACS for \Cii\ spectroscopy (large red
     circles) and the full parent sample of 84 sources (black empty
     circles). Our targets cover approximately two orders of magnitude
     in \LIR. We overplot the Revised Imperial {\it IRAS}--FSC Redshift
     Catalogue of \citet{Wang14} [blue crosses; considering
       $L(8-1000\mu{\rm m})/L_\odot = 4\pi D_{\rm L}^2 \times S_{\rm
         IR}$, where $S_{\rm IR}=1.8\times 10^{-14}\,(13.48\,S_{12{\rm
           \mu m}} + 5.16\,S_{25{\rm \mu m}} + 2.58\,S_{60{\rm \mu m}}
       + S_{100{\rm \mu m}})$\,W\,m$^{-2}$; \citealt{Sanders96}], and
     the observed {\it H}-ATLAS sources in the GAMA-09h field (green
     crosses). The dotted line shows the detection threshold used to
     select our targets, above which sources were randomly selected
     (any objects lying below the threshold are due to an improved SED
     fitting approach that was implemented post-selection, resulting
     in slightly lower \LIR\ estimates). The detection threshold was
     applied to ensure a $>$\,$5\sigma$ line detection within 10\,min
     using the simplest (single scan, single pointing) PACS
     spectroscopic mode, i.e.\ $S_{\rm
       [CII]}\,>\,15.5\times10^{-18}$\,W\,m$^{-2}$ assuming \Cii/IR =
     0.001.}
   \label{fig_threshold}
\end{figure}

\section{Sample selection for PACS spectroscopy}

\begin{figure*}
   \centering
   \includegraphics[scale=0.88]{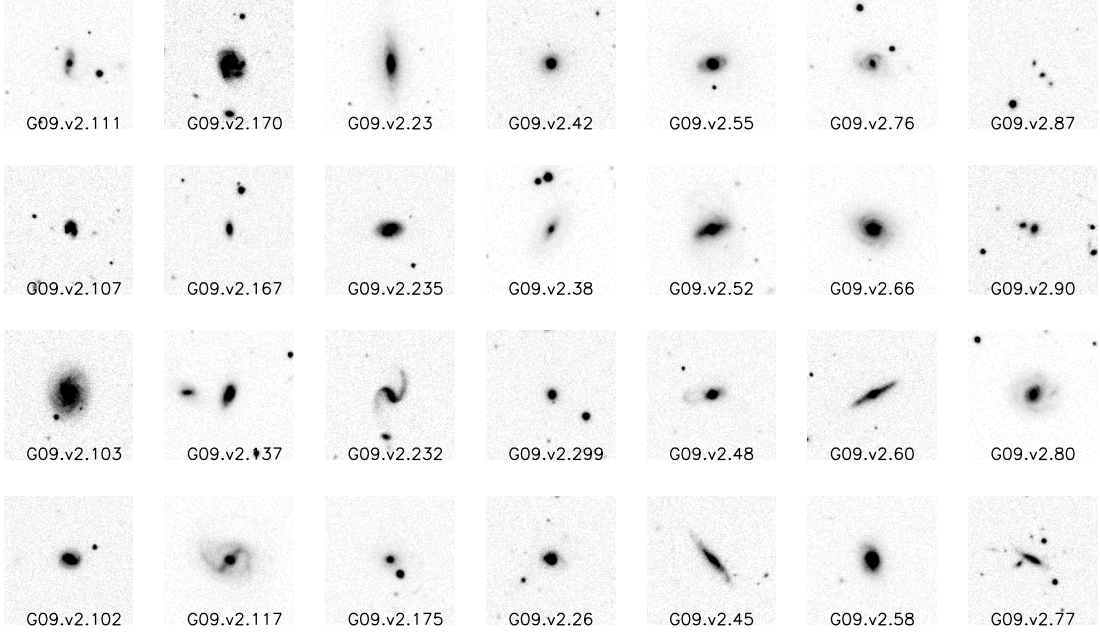}
   \caption{Postage-stamp images ($1'\times1'$) for our targets taken from the
     SDSS $r$-band imaging.}
   \label{fig_postages}
\end{figure*}

We make use of the internal phase-1 data release (v2;
\citealt{Rigby11, Smith11}, Valiante et al.\ in prep., and Bourne et
al.\ in prep) taken from the three equatorial fields of the {\it
  Herschel}-Astrophysical Terahertz Large Area Survey \citep[{\it
    H}-ATLAS\footnote{{\sc www.h-atlas.org}}--][]{Eales10}. We make
use of both Spectral and Photometric Imaging Receiver (SPIRE;
\citealt{Griffin10}) and PACS {\it H}-ATLAS data
(\citealt{Pascale11,Ibar10}). From the total number of 109,231 sources
detected in these fields, we performed the following selection
criteria for our targets: (1) a flux density threshold at $S_{\rm
  160\mu m}>150\,$mJy, where 160\,$\mu$m is near the peak of the
spectral energy distribution (SED) of a local star-forming galaxy; (2)
targets without PACS 160-$\mu$m ($>3$\,$\sigma$) neighbours within
2\,arcmin from their centroids (to avoid problems when chopping; see
\S\,\ref{sect2_obs}); (3) an unambiguously identification in the Sloan
Digital Sky Survey \citep[SDSS DR6 --][]{Adelman-McCarthy08} ({\sc
  reliability}\,$>0.8$, \citealt{Smith11}, Bourne et al.\ in prep.);
(4) galaxies need to be smaller than the PACS spectroscopic field of
view, so restricted to sources with Petrosian SDSS radii smaller than
15\,arcsec in the $r$-band (see Fig.~\ref{fig_postages}); (5)
high-quality spectroscopic redshifts from the Galaxy and Mass Assembly
survey \citep[GAMA\footnote{\sc www.gama-survey.org} --][]{Driver09,
  Driver11} ({\sc z\_qual}\,$\ge$\,3). Note that GAMA combines spectra
from SDSS with deeper spectra taken with the AAOmega fibre-fed
spectrograph on the 3.9-m Anglo-Australian Telescope; (6) a redshift
between $0.02<z<0.2$ (median, 0.05), where the upper limit is imposed
by the point at which the \Cii\ emission is redshifted to the edge of
the PACS spectrometer 160\,$\mu$m band (which is
known to leak), and the lower limit is imposed by the Petrosian
$r$-band criterion. Applying all these criteria, we remain with a
sample of 327 sources.

For each field, we randomly selected galaxies to span a wide range of
optical morphological types and IR luminosities. In total, we selected
a statistically significant sample of 84 galaxies for a
\Cii\ spectroscopic campaign using {\it Herschel}-PACS during its
second open time call. Galaxies have photometric detections at 100,
160, 250, 350 and 500\,$\mu$m by {\it Herschel}, including Wide-field
Infrared Survey Explorer \citep[{\it WISE} --][]{Wright10} photometry,
and approximately half of them were also detected by {\it IRAS} at
60\,$\mu$m (using the Revised Imperial {\it IRAS}--FSC Redshift
Catalogue\footnote{\sc astro.ic.ac.uk/home/mrrobinson}, {\sc
  rifscz\_short.v2}; \citealt{Wang14}). The combination of {\it
  IRAS}/{\it WISE}/{\it Herschel}/SDSS photometry and GAMA
spectroscopy allows accurate estimates for a wide range of physical
parameters, including \LIR\ (having a range of $10^{10}<L_{\rm
  IR}/L_\odot<10^{12}$), dust temperature, stellar masses, emission
line strengths, metallicities, etc.\ (see \S\,\ref{results_section}).

\section{PACS IFU observations and data analysis}
\label{sect2_obs}

Data were obtained between 2013 April 18--28, a few weeks before {\it
  Herschel}\ ran out of liquid helium, for the {\it Herschel} Open
Time project, {\sc ot2\_eibar\_1}
(P.I.\ E.\ Ibar). Of the parent sample of 84
galaxies, 28 were observed (Table~\ref{table_obsids}), all from only
one of the equatorial {\it H}-ATLAS coverages -- the GAMA-09\,h
field. The selection of these 28 was performed purely on the basis of
scheduling efficiency, so this sample is representative of the parent
sample, just smaller in number (see Fig.~\ref{fig_threshold}).

\begin{table}
  {\scriptsize{
  \label{table_obsids}
  \caption{Log of PACS spectroscopic observations.
      Each observation lasted 394\,s and was processed with SPG
      version 12.1.0. The table shows the IAU name, \hatlas's
      nickname, GAMA identifier and the {\sc obsid} during the {\it
        Herschel} campaign.}  \centering
  \begin{tabular}{lccc}
    \hline
    IAU name &
    Nickname &
    GAMA ID &
    {\sc obsid} \\
    \hline
    HATLAS\,J09:17:21.91$+$00:19:18.8 & G09.v2.117	& 601323 & 1342271048 \\
    HATLAS\,J09:12:05.82$+$00:26:55.5 & G09.v2.42	& 216401 & 1342270763 \\
    HATLAS\,J09:09:49.59$+$01:48:45.9 & G09.v2.26	& 324842 & 1342270761 \\
    HATLAS\,J09:07:50.13$+$01:01:42.6 & G09.v2.107	& 279387 & 1342270762 \\
    HATLAS\,J09:05:32.66$+$02:02:20.0 & G09.v2.58	& 382362 & 1342270757 \\
    HATLAS\,J09:00:04.98$+$00:04:46.7 & G09.v2.55	& 209807 & 1342270755 \\
    HATLAS\,J08:58:35.96$+$01:31:48.9 & G09.v2.76	& 376679 & 1342270657 \\
    HATLAS\,J08:58:28.62$+$00:38:14.8 & G09.v2.80	& 622694 & 1342270756 \\
    HATLAS\,J08:57:48.00$+$00:46:41.2 & G09.v2.48	& 622662 & 1342270656 \\
    HATLAS\,J08:54:50.33$+$02:12:08.9 & G09.v2.38	& 386720 & 1342270658 \\
    HATLAS\,J08:54:06.05$+$01:11:30.5 & G09.v2.137	& 301346 & 1342270655 \\
    HATLAS\,J08:53:56.59$+$00:12:55.6 & G09.v2.170	& 600026 & 1342270649 \\
    HATLAS\,J08:53:46.47$+$00:12:51.6 & G09.v2.45	& 600024 & 1342270648 \\
    HATLAS\,J08:53:40.87$+$01:33:48.1 & G09.v2.103	& 323855 & 1342270654 \\
    HATLAS\,J08:52:34.39$+$01:34:19.8 & G09.v2.87	& 323772 & 1342270653 \\
    HATLAS\,J08:51:12.83$+$01:03:43.6 & G09.v2.235	& 371789 & 1342270651 \\
    HATLAS\,J08:51:11.48$+$01:30:06.9 & G09.v2.60	& 376293 & 1342270652 \\
    HATLAS\,J08:49:07.15$-$00:51:40.2 & G09.v2.175	& 3624571 & 1342270647 \\
    HATLAS\,J08:46:30.79$+$00:50:55.1 & G09.v2.90	& 278475 & 1342270650 \\
    HATLAS\,J08:44:28.41$+$02:03:49.8 & G09.v2.52	& 345754 & 1342270371 \\
    HATLAS\,J08:44:28.27$+$02:06:57.3 & G09.v2.77	& 386263 & 1342270370 \\
    HATLAS\,J08:43:50.90$+$00:55:34.0 & G09.v2.102	& 371334 & 1342270372 \\
    HATLAS\,J08:43:05.18$+$01:08:57.0 & G09.v2.167	& 300757 & 1342270373 \\
    HATLAS\,J08:42:17.71$+$02:12:22.3 & G09.v2.232	& 867786 & 1342270365 \\
    HATLAS\,J08:41:39.45$+$01:53:44.8 & G09.v2.299	& 345647 & 1342270364 \\
    HATLAS\,J08:38:32.01$+$00:00:44.5 & G09.v2.111	& 208589 & 1342270374 \\
    HATLAS\,J08:37:45.33$-$00:51:42.3 & G09.v2.66	& 3895257 & 1342270362 \\
    HATLAS\,J08:36:01.57$+$00:26:18.1 & G09.v2.23	& 214184 & 1342270363 \\
    \hline
  \end{tabular}
  }}
\end{table}

\begin{figure*}
   \centering
   \includegraphics[scale=0.85]{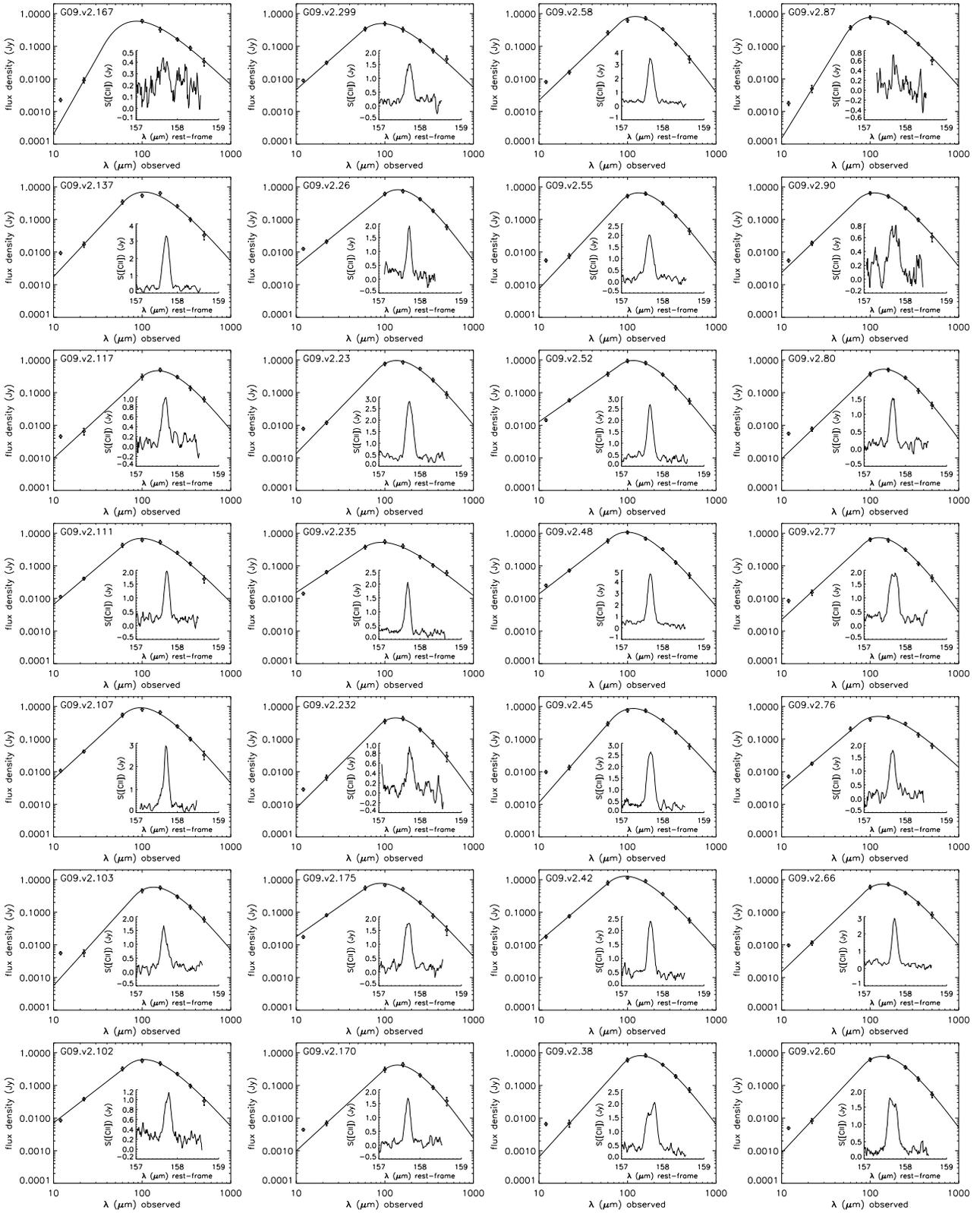}
   \caption{The flux densities versus observed wavelengths for all 28
     galaxies presented in this work. From short to long wavelengths,
     SED data points correspond to broadband photometry from: {\it
       WISE} 12 and 22\,$\mu$m, {\it IRAS} 60\,$\mu$m,
     {\it Herschel} PACS 100, 160, and {\it Herschel}
       SPIRE 250, 350, and 500\,$\mu$m. Note that {\it WISE} 12 is not
       used to fit the SED. All photometric points have significance
       above 3-$\sigma$ limit. The {\it inset} in each panel shows the
       \Cii\ spectra, flux density versus rest-frame wavelength,
       observed by the central spaxel of the PACS detector. Only two
       galaxies (G09.v2.87 and G09.v2.167) resulted with a \Cii\ peak
       to noise line ratio lower than three. These two galaxies appear
       as upper limits in the following figures.}
   \label{fig_SEDs}
\end{figure*}

\begin{figure*}
   \centering
   \includegraphics[scale=0.8]{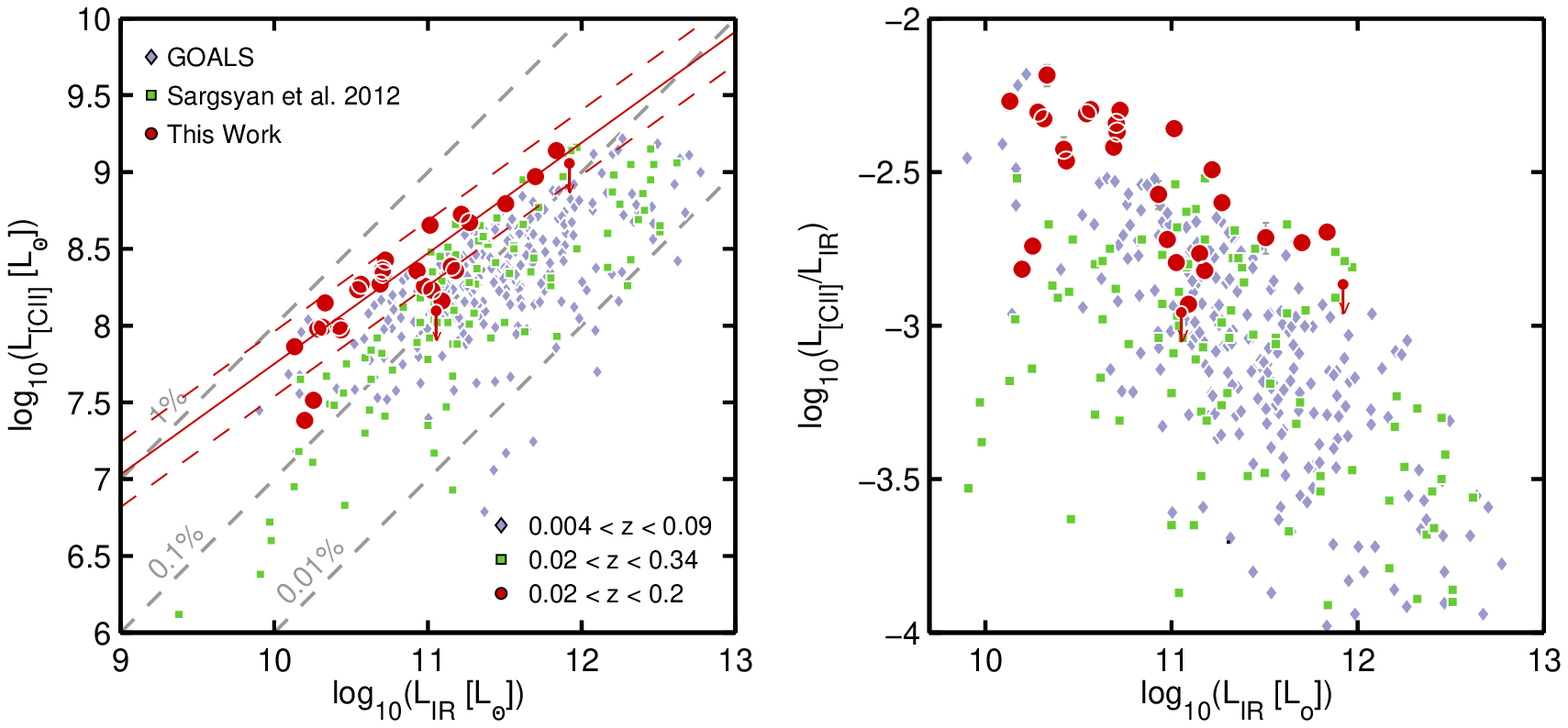}
   \caption{{\it Left:} \Cii\ line luminosity versus bolometric IR
     luminosity for the sample of sources presented in this work. We
     compare our observations to previous (at similar redshifts),
     GOALS \citep{Diaz-Santos13} and \citet{Sargsyan12}'s
     samples. \citeauthor{Diaz-Santos13}'s data have been scaled by a
     1.75 factor to convert from far-IR to IR luminosities (see
     Appendix E from \citealt{Herrera-Camus15}). Continuum and dashed
     red lines show the best linear fit (excluding \Cii\ undetections)
     ${\rm log_{10}}(L_{\rm [CII]}/L_\odot) = (0.721\pm0.003)\times
     {\rm log}_{10}(L_{\rm IR}/L_\odot)+(0.53\pm0.04)$ and the 0.2\,dex
     scatter, respectively. Diagonal lines show the 1.0, 0.1 and 0.01
     per cent fraction of \LIR. {\it Right:} \Cii/\LIR as a function
     of \LIR, revealing the well-known \Cii\, deficit. Colour coded
     points are the same in both figures.}
   \label{fig_CII_deficit}
\end{figure*}

We targeted the redshifted \Cii-157.7\,$\mu$m line emission using PACS
spectroscopy in the first order/{\sc r1} filter. PACS comprises an
array of $5\times 5$ spaxels, 9.4\,arcsec $\times$ 9.4\,arcsec each,
covering $\sim$\,2$\,\mu$m of bandwidth covered. The
instantaneous field of view in a single pointing (pointed-mode)
observation is thus 47\,arcsec $\times$ 47\,arcsec, hence based on the
optical size, the majority of the line emission falls on the central
spaxel.

The observations were made with a small chopping angle (1.5\,arcmin),
as appropriate for our sources. Note that our
selection criteria have ensured that no sources would lie in the
chopped beam. The {\it H}-ATLAS survey targeted areas of low IR
background, so that the background intensities never approached the
detector saturation level.

We retrieved the calibrated PACS level-2 data
  products (processed with SPG v12.1.0) using the {\it Herschel} User
  Interface\footnote{\sc archives.esac.esa.int/hsa/ui/hui.jnlp} ({\sc
    HUI v6.0.4}). We exported the {\sc hps3drb} (blue) and {\sc
    hps3drr} (red) data cubes, including the appended signal,
  coverage, noise and wavelength index. These data cubes comprise $5
  \times 5$ spaxels $\times\, n_{\rm chan}$ rebinned spectral channels
  for each observed galaxy. At the observed frequencies, the effective
  spectral resolution is $\sim$\,190--240\,km\,s$^{-1}$, providing
  useful kinematical information for typical disk-like galaxies (see
  Table\,\ref{table1}). We use the Interactive Data Language ({\sc
    idl}) to analyse these cubes along with the wavelength index
  array.

First, we find the best flux density measurement by comparing the
central spaxel line emission (plus an aperture correction
$\sim$\,0.4--0.5; a point-like estimate) with the summed over the
whole IFU (extended estimate). We conclude that the best compromise in
terms of signal-to-noise for the line flux density is to use the
weighted sum (aided by the appended instrumental noise cube) of the
central $3\times3$\,spaxels. The addition of the outer spaxels only
introduced noise in the signal. We recover, on average, 10\% more flux
in the extended estimate than in the point-like one.

\Cii\ line fluxes were measured via a Gaussian fit to the added
spectra (see Fig.~\ref{fig_SEDs}), fitting simultaneously a linear
background slope and a Gaussian using the {\sc mpfitpeak} routine
within {\sc idl}. To perform the fit, we removed channels suffering
from higher noise in both ends of the spectrum.

Using the spectroscopic redshifts, we calculate
  \Cii\ luminosities following \citet{Solomon05}. In order to estimate
  uncertainties for the line parameters, we ran a Monte-Carlo
  realisation ($1000\times$), randomly varying the signal per spectral
  channel using the instrumental error cube. Based on these simulated
  data, we quote 1-$\sigma$ uncertainties for the line measurements
  and 3-$\sigma$ upper limits (based on intrinsic 400\,km\,s$^{-1}$
  FWHM widths) in Table~\ref{table1}.

We opted to receive data in order/{\sc b3} (blue filter at 70\,$\mu$m)
rather than order/{\sc g2} (green), as we already have 100-$\mu$m
continuum photometry from the {\it H}-ATLAS imaging survey. This PACS
blue point was expected to be used to improve the photometric SED
sampling of the targets. We explored the blue spectra in the
53--63\,$\mu$m range (depending on the source redshift) and found no
clear signs of line emission in the data cubes, as expected. The
significance of the continuum level (blue and red) was not high enough
to permit a reliable photometric point for our SED fitting approach.

\section{Results}
\label{results_section}

\begin{figure*}
   \centering
   \includegraphics[scale=0.55]{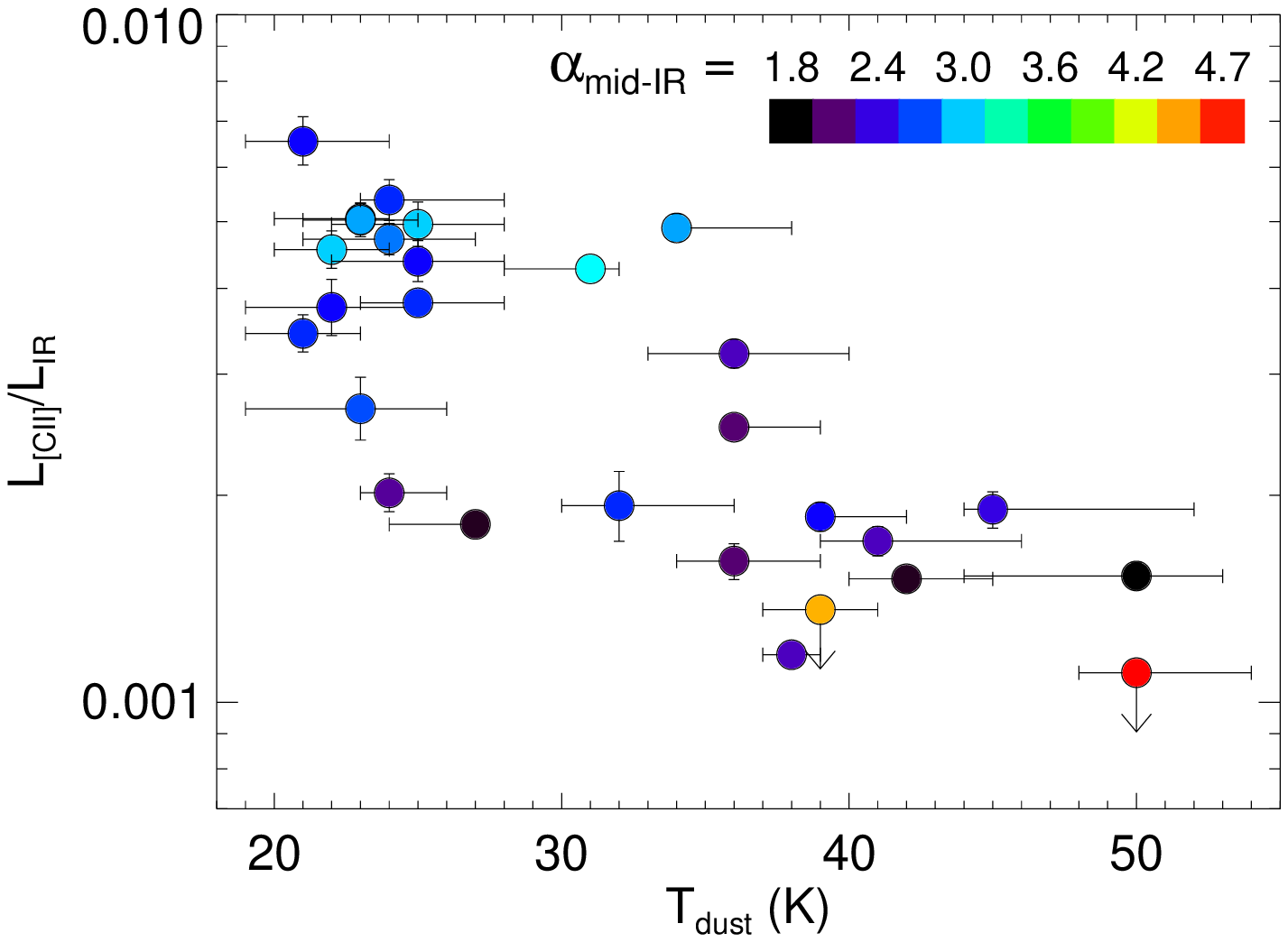}
   \includegraphics[scale=0.55]{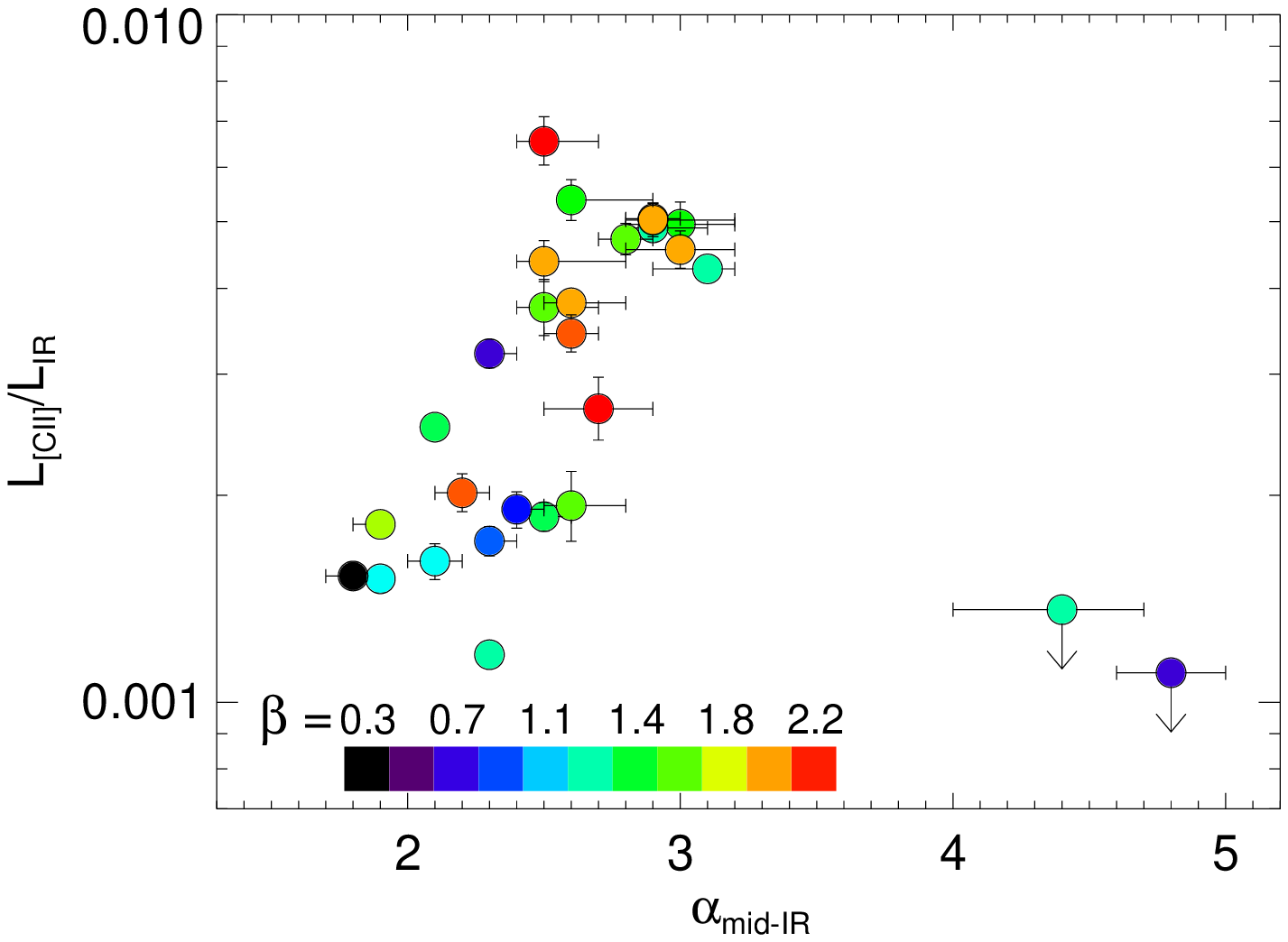}
   \includegraphics[scale=0.55]{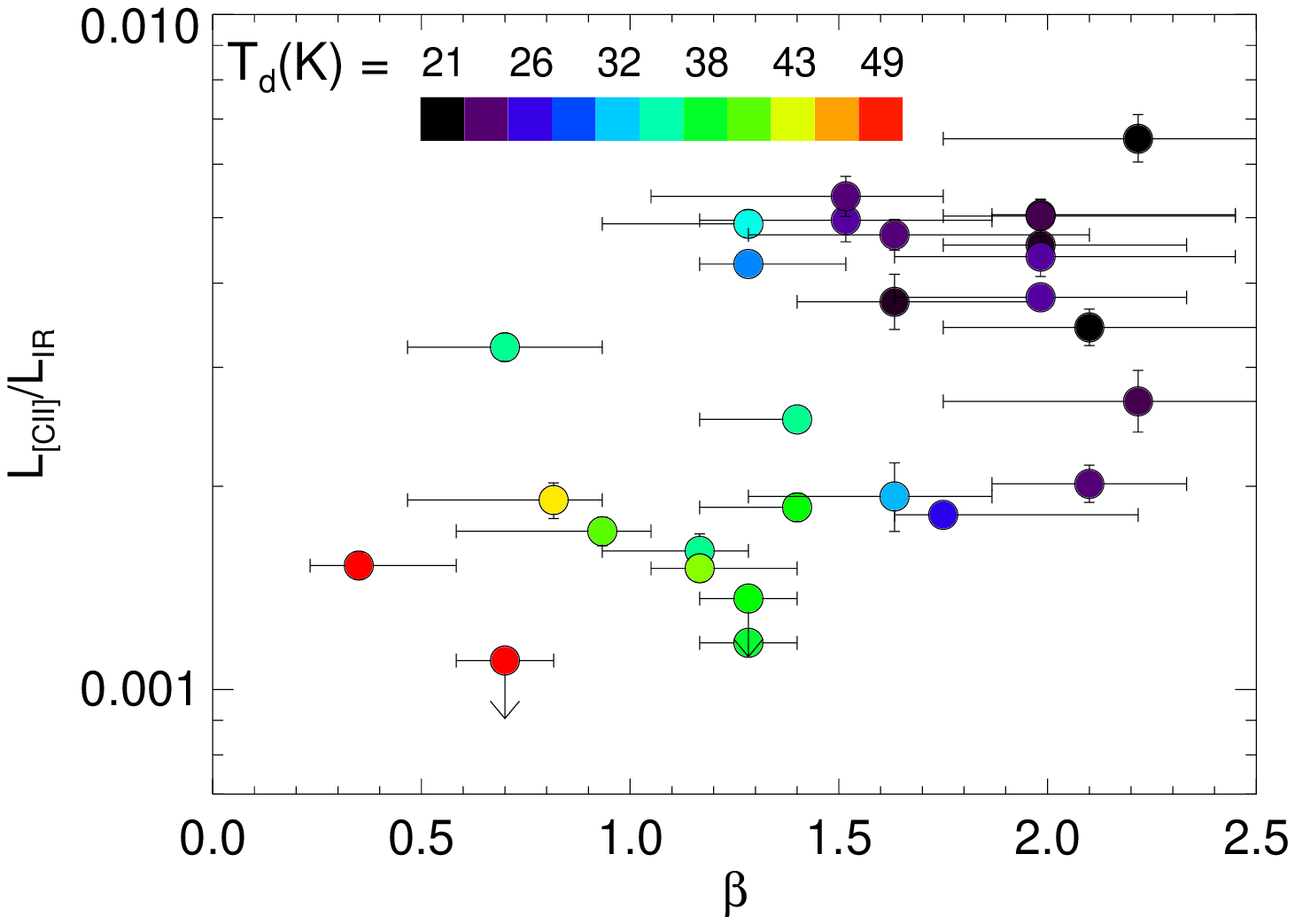}
   \caption{Dependency of the \Cii/IR luminosity ratio as a function
     of dust temperature (top-left; colour-coded in mid-IR index), mid-IR
     index (top-right; colour-coded in $\beta$) and dust emissivity
     $\beta$ index (bottom; colour-coded in dust temperature). Data
     suggest that galaxies with colder dust temperatures tend
     to have higher \Cii/IR ratios than hotter galaxies.}
   \label{fig_Tdust}
\end{figure*}

\subsection{SED Properties}

Making use of the {\it Herschel}-PACS, {\it Herschel}-SPIRE, and the
publicly available broadband measurements from {\it WISE}-22\,$\mu$m
(all targets detected) and {\it IRAS} (17 sources reliably detected at
60\,$\mu$m), we fitted the SED of each galaxy following a similar
approach as in \citet{Ibar13}. We fit a usual modified black body but
forcing the SED shape to a power law in the high-frequency range of
the spectra. The fits are aided by the known optical redshifts. We
shift all SEDs to rest-frame frequencies, following ($\nu=\nu_{\rm
  obs}\,[1+z]$), and then fit the following parametrisation

\begin{equation}
  S_\nu(\nu) = 
  A \times \left\{ 
  \begin{array}{ll}
    {\rm MBB}(\nu) & 
    {\rm if}\quad \nu\le\nu_\star      \\
    {\rm MBB}(\nu_\star)\times(\nu/\nu_\star)^{-\alpha_{\operatorname{mid-IR}}} & 
    {\rm if}\quad \nu>\nu_\star \\
  \end{array} 
  \right. 
\end{equation}

\noindent
where

\begin{equation}
  {\rm MBB}(\nu) = \frac{\nu^{3+\beta}}
  {{\rm exp}\left(\frac{h\,\nu}{k\,T_{\rm d}}\right)-1}.
\label{mbb_eqn}
\end{equation}

The $h$ and $k$ parameters refer to Planck and Boltzmann constants,
respectively. $\beta$ is the usually called dust emissivity index
\citep[e.g.][]{Seki80,Dunne11} -- an averaged property of the dust grain
emission over of whole galaxy. The parameter $\nu_\star$ is
obtained numerically at

\begin{equation}
\frac{d\,{\rm log_{10}(MBB)}}
{d\,{\rm log}_{10}(\nu)}(\nu_\star) =\alpha_{\operatorname{mid-IR}}
\end{equation}

\noindent
which is simply used to match the slope of the modified black body
function with the high-frequency power-law (roughly at
$\sim$100--200\,$\mu$m). This parameter does not have a physical
meaning, nevertheless it is useful to account for the the mid-IR part
of the spectra which otherwise is underestimated by a simple modified
black-body (see Appendix~\ref{AppendixA}). Examples of best fit SED
are shown in Fig.~\ref{fig_SEDs}. To measure the IR luminosity,
$L(8-1000\mu {\rm m})$, we integrate the best fitted SED in rest-frame
frequencies between $\nu_1=0.3$\,THz (1000\,$\mu$m) and
$\nu_2=37.5$\,THz (8\,$\mu$m),

\begin{equation}
  L(8-1000\,\mu{\rm m}) =
  4\pi\,D_{\rm L}^2(z)\,\int_{\nu_{\rm 1}}^{\nu_{\rm 2}}S_\nu\,d\nu
\end{equation}

Under this parametrisation, we obtain the dust temperature ($T_{\rm
  d}$), the dust emissivity index ($\beta$), the mid-IR slope
($\alpha_{\operatorname{mid-IR}}$), and the normalisation which
provides the total IR luminosity (rest-frame
8--1000\,$\mu$m). Uncertainties for each parameter were obtained from
a Monte-Carlo simulation ($100\times$), randomly varying the broadband
photometry as appropriate for their measured uncertainties
(Table~\ref{table1}). The best-fit SEDs for all our target galaxies
are shown in Fig.~\ref{fig_SEDs}. We find that our sample have
luminosities of the order of $10<{\rm log_{10}(L_{\rm IR}/{\rm
    L}_\odot)}<12$, dust temperatures of $20<T_{\rm d}/{\rm K}<55$,
mid-IR slopes in the range of $1.5<\alpha_{\operatorname{mid-IR}}<3.2$
(plus two outliers at $\sim$\,4.5), and dust emissivity indices of
$0.3<\beta<2.5$.

\subsection{The \Cii\ deficit}

Like many before us, we find that \Cii\ luminosity (including PDR and
non-PDR components, which cannot be disentangled in this study)
correlates strongly with \LIR\ luminosity over approximately two
orders of magnitude (Fig.~\ref{fig_CII_deficit}). A
  simple linear regression (excluding \Cii\ undetections) results in ${\rm log_{10}}(L_{\rm
    [CII]}/L_\odot) = (0.721\pm0.003)\times {\rm log}_{10}(L_{\rm
    IR}/L_\odot)+(0.53\pm0.04)$ and the distribution has a Spearman's
  rho rank correlation coefficient of 0.88 and a two-sided significance
  of its deviation from cero of $2\times10^{-7}$. The slope of this
correlation evidences the so-called \Cii\ deficit, i.e.\ at higher IR
luminosities the line to continuum ratio decreases.

\begin{figure}
   \centering
   \includegraphics[scale=0.48]{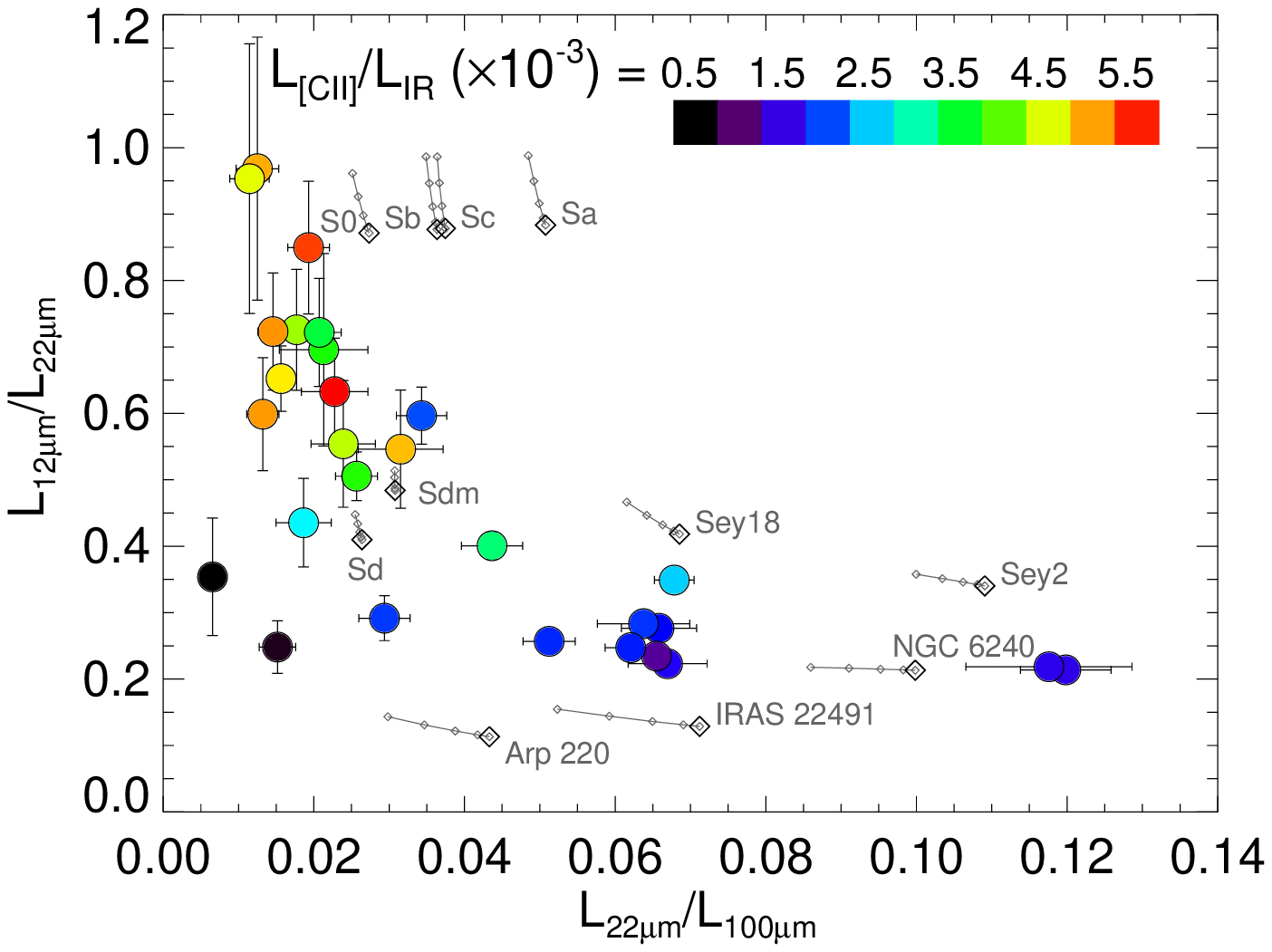}
   \includegraphics[scale=0.48]{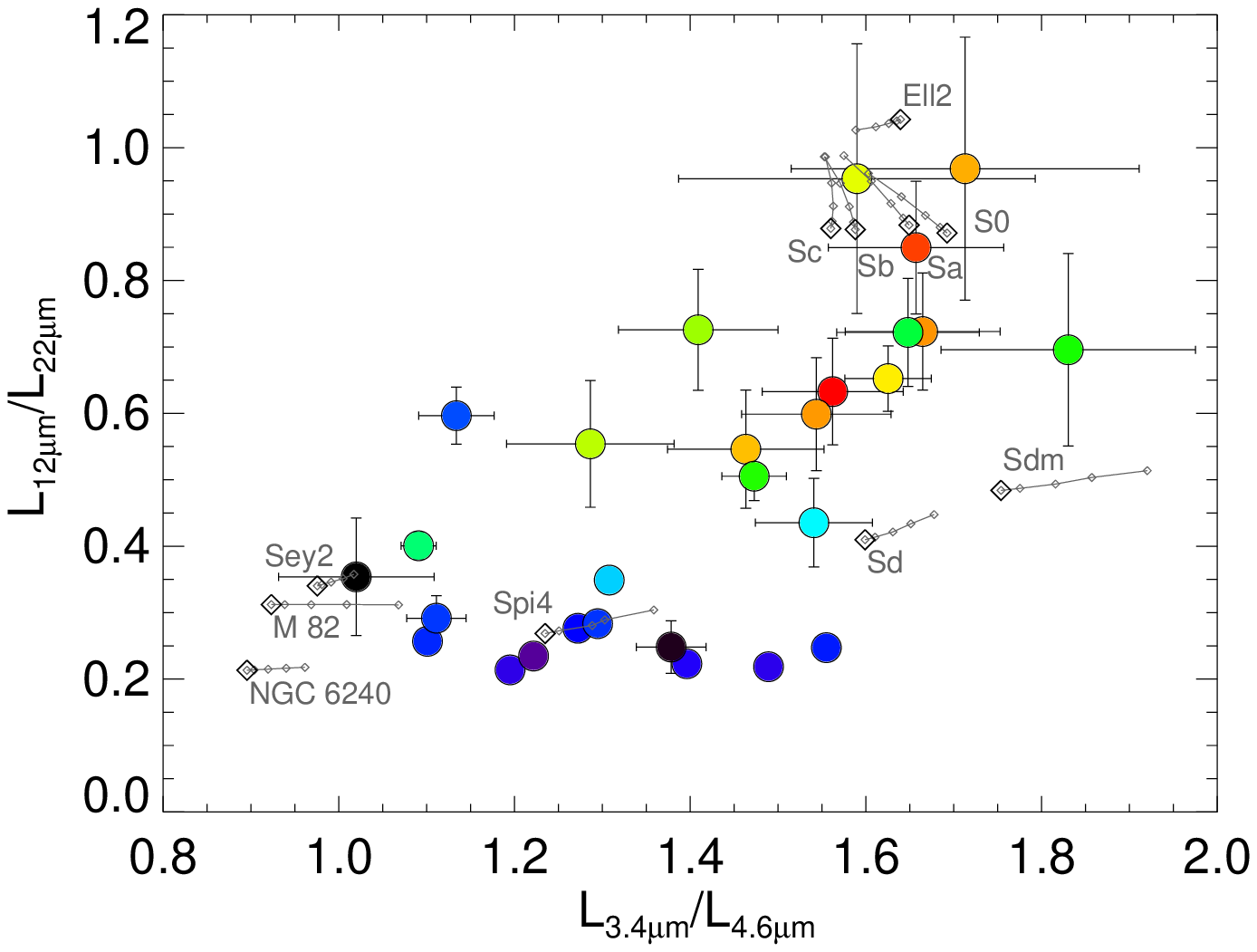}
   \caption{The figure shows the locus for our galaxies in a couple of
     colour-colour plots, exploiting {\it Herschel} and {\it WISE}
     photometry. Given the optical extension of the targets, we
     consider elliptical aperture photometry for the {\it WISE}
     photometry. Broad-band luminosities have not been
     $k$-corrected. Data points are colour-coded as a function of the
     \Cii/IR luminosity ratio (see top-bar in left panel). We overplot
     also the expected colours for different SED templates taken from
     \citet{Polletta07}'s libraries (convolved with response filters)
     in five redshift ranges ($z=$\,0, 0.05, 0.1, 0.15 and 0.2)
     indicated by small diamonds, where the biggest one is at
     $z=0$. The figure indicates that galaxies with highest \Cii/IR
     ratios are those with $0.5<S_{\rm 12\mu m}/S_{\rm 22\mu m}<1.0$,
     galaxies which could be explained by normal spiral `{\it
       Hubble}-class' galaxies (e.g.\ S0, Sa, Sb, Sc), while those
     with lower values are better represented by star-burst templates
     such as Arp\,220, M\,82, IRAS\,22491, NGC\,6240, etc.}
   \label{deficit_WISE}
\end{figure}

\subsection{Correlations with dust properties}

The quality of our data permits us to explore dust temperatures in
great detail (uncertainties range from 1 to $\sim$\,5\,K based on
Monte-Carlo simulations; see Table~\ref{table1}). This is a major step
forward compared to previous studies that only
  provided a proxy for the dust temperature, e.g.\ using the {\it
  IRAS} 100\,$\mu$m to 60\,$\mu$m ratio (\citealt{Malhotra01}). Within
our sample, most of the low-luminosity, low-redshift galaxies with
\LIR\,$\lesssim10^{11}\,$L$_\odot$, have isothermal dust temperatures
of $T_{\rm d}=23\pm1$\,K (Fig.~\ref{fig_Tdust}) -- typical of `normal'
$H$-ATLAS galaxies (e.g.\ \citealt{Smith13}). The galaxies with
cooler dust ($T_{\rm d}<30$\,K) tend to have $\sim\,2\,\times$ higher
\Cii/IR ratios than warmer galaxies. We find that $\sim$\,90 per cent
of the galaxies with $T_{\rm d}<30$\,K have
\Cii/IR\,$>3\times10^{-3}$. This fraction decreases to $\sim$\,20 per
cent for hotter galaxies.

It is interesting to note the mild correlation where flatter mid-IR
slopes tend to produce lower \Cii/IR ratios. This might be a
manifestation for the presence of stronger radiation fields
contributing to the 22\,$\mu$m flux density, or simply due to the poor
mid-IR constraints on the SED-fitting approach (see also
\S\,\ref{subsec_wise}). We find that the fitted dust
  emissivity index, $\beta$, correlates with the \Cii/IR ratio, where
  higher $\beta$ values are found in galaxies with higher \Cii/IR
  ratios (Fig.~\ref{fig_Tdust}). In Table~\ref{table_correlations} we
explore various possible correlations for the \Cii/IR ratio as a
function of fitted and observed quantities.

\subsection{Correlations with {\it WISE} colours}
\label{subsec_wise}

Fig.~\ref{deficit_WISE} shows a colour-colour diagram using {\it WISE}
and {\it Herschel} photometry for our sample. We find
that low \Cii/IR ratios ($<$\,2.5\,$\times10^{-3}$) are almost
ubiquitous among galaxies presenting a flux density ratio $S_{\rm
  12\mu m}/S_{\rm 22\mu m}<0.5$, i.e.\ a power-law with
$\alpha^{12}_{22}<-1.14$. Note that all of our galaxies got fitted
$\alpha_{\operatorname{mid-IR}}$ values which are steeper than $-1.14$
(see Fig.~\ref{fig_Tdust}), implying that the {\it WISE} colours do
not reflect the same mid-IR slope. In fact, Fig.~\ref{deficit_WISE}
indicates that the \Cii/IR luminosity ratio depends strongly as a
function of the 22\,$\mu$m flux density (see also
Table~\ref{table_correlations}).

In Fig.~\ref{deficit_WISE}, we also show the expected {\it WISE}
and {\it Herschel} colours from different SED
templates taken from \citet{Polletta07} and convolved with {\it WISE}
transmission filters in the redshift range of this study ($z<0.2$). We
find that mid-IR templates from star-bursting galaxies
(e.g.\ NGC\,6240) and moderately luminous AGNs (e.g.\ Sey2) are well
suited to explain low $S_{\rm 12\mu m}/S_{\rm 22\mu m}$ ratios
(i.e.\ lower \Cii/IR ratios). Higher \Cii/IR ratios ratios tend to be
better explained by spiral {\it Hubble}-type galaxies (S0--Sc). A
close inspection to their mid-IR to far-IR SEDs, suggests three
signatures controlling the $S_{\rm 12\mu m}/S_{\rm 22\mu m}$ ratio;
the strength of the PAH 7.7\,$\mu$m emission line system, the Silicate
absorption band at 9.8\,$\mu$m and the slope of the
spectra. Unfortunately, with the present data we are unable to
distinguish between these different features. We have performed a
basic simulation taking into account the broadband filters and
artificial SEDs including different slopes, equivalent widths of the
PAH emission and Si absorption band. We find that the mid-IR slope is
probably the dominant component in varying the observed $S_{\rm 12\mu
  m}/S_{\rm 22\mu m}$ ratio. On the other hand, on average, galaxies
with low (high) \Cii/IR ratios have low (high) $S_{\rm 3.4\mu
  m}/S_{\rm 4.6\mu m}\approx1.3$ (1.45) ratios. Based on
\citeauthor{Polletta07}'s templates, we find that {\it WISE} colours
from elliptical, or normal early-type galaxies, or powerful AGNs, lie
outside Fig.~\ref{deficit_WISE}, in agreement with the far-IR
selection criteria which prefers star-forming dusty galaxies.

\begin{landscape}
\begin{table}
\centering
\scriptsize{
\caption{The correlation of the $L_{\rm [CII]}/L_{\rm IR}$ luminosity
  ratio as a function of various different parameters. Linear
  regression values, where $L_{\rm [CII]}/L_{\rm IR} = a + b\times
  \psi$, are obtained by simple $\chi^2$ minimisation. We exclude the
  two \Cii\ undetections from this analysis. Last two columns show the
  Spearman's rho ($\rho$) rank correlation and the two-sided
  significance of its deviation from cero ($Q$). We provide values for
  the best linear fit, only when a significant correlation is found
  $Q<5$\,\%.}\label{table_correlations}
\begin{tabular}{ccccc}
\hline
$\psi$ &
$a$ &
$b$ &
$\rho$ &
$Q(\%)$ \\
\hline
${\rm log_{10}}[L_{\rm IR} (L_\odot)]$ & -0.7121$\pm$ 0.1021 & -0.1713$\pm$ 0.0094 & -0.49 &  1.1 \\
$T_{\rm dust}$ (K) & -1.9803$\pm$ 0.0180 & -0.0182$\pm$ 0.0005 & -0.72 &  0.0 \\
$\alpha_{\operatorname{mid-IR}}$ & -3.5550$\pm$ 0.0256 &  0.4082$\pm$ 0.0105 &  0.75 &  0.0 \\
$\beta$ & -2.8705$\pm$ 0.0141 &  0.2084$\pm$ 0.0094 &  0.52 &  0.7 \\
${\rm log_{10}}[L_{\rm 22\mu m}({\rm erg/s/Hz})]$ &  4.9372$\pm$ 0.2326 & -0.2388$\pm$ 0.0074 & -0.76 &  0.0 \\
${\rm log_{10}}[L_{\rm 100\mu m}({\rm erg/s/Hz})]$ &  1.4705$\pm$ 0.3232 & -0.1231$\pm$ 0.0098 & -0.42 &  3.3 \\
${\rm log_{10}}[L_{\rm 160\mu m}({\rm erg/s/Hz})]$ & -- & -- & -0.31 & 12.8 \\
${\rm log_{10}}[L_{\rm 250\mu m}({\rm erg/s/Hz})]$ & -- & -- & -0.26 & 19.2 \\
${\rm log_{10}}[L_{\rm 350\mu m}({\rm erg/s/Hz})]$ & -- & -- & -0.26 & 19.3 \\
${\rm log_{10}}[L_{\rm 500\mu m}({\rm erg/s/Hz})]$ & -- & -- & -0.25 & 21.4 \\
$L_{\rm 3.4\mu m}/L_{\rm 4.6\mu m}$ & -3.4390$\pm$ 0.0329 &  0.6224$\pm$ 0.0234 &  0.62 &  0.1 \\
$L_{\rm 12\mu m}/L_{\rm 22\mu m}$ & -2.9487$\pm$ 0.0096 &  0.8443$\pm$ 0.0194 &  0.87 &  0.0 \\
$L_{\rm 22\mu m}/L_{\rm 100\mu m}$ & -2.2960$\pm$ 0.0079 & -5.5447$\pm$ 0.1360 & -0.82 &  0.0 \\
$r_{\rm eff,50\%}({\rm kpc})$ & -- & -- &  0.19 & 36.4 \\
$r_{\rm eff,90\%}({\rm kpc})$ & -- & -- & -0.04 & 82.8 \\
${\rm log_{10}[\Sigma_{\rm IR}(L_\odot/{\rm kpc}^2)]}$ & -0.3570$\pm$ 0.0792 & -0.2462$\pm$ 0.0088 & -0.65 &  0.0 \\
\hline
\end{tabular}
}
\end{table}

\begin{table}
  \caption{
      Properties of the targets analysed in this work. 
    {\it Col1:} {\it H-}ATLAS's nickname;
    {\it Col2:} GAMA's redshift;
    {\it Col3:} Line peak flux;
    {\it Col4:} Intrinsic line FWHM width obtained after removing (in quadrature) the intrumental spectral resolution;
    {\it Col5:} Line flux density ($\sqrt{2\pi}\times P_{\rm [CII]}\times$fitted line width);
    {\it Col6:} Line luminosity, where upper limits are $3-\sigma$ assuming an intrinsic 400\,km\,s$^{-1}$ FWHM width; 
    {\it Col7:} IR (8-1000$\mu$m) luminosity;
    {\it Col8:} Dust temperature;
    {\it Col9:} Dust emissivity index;
    {\it Col10:} Mid-IR slope;
    {\it Col11:} $L_{[{\rm C\,{\sc II}}]}$/\LIR\ luminosity ratio, where upper limits use values from Column~6;
    {\it Col12:} Effective radius at which 50\% of the power is encircled using SDSS $r$-band imaging;
    {\it Col13:} Stellar mass;
    {\it Col14:} Specific star-formation rate using the IR luminosity as proxy;
    {\it Col15:} Metallicity measurements for star-forming galaxies;
    {\it Col16:} BPT classification, 0$=$star-forming, 1$=$composite and 2$=$AGN;
    {\it Col17:} Optically based visual morphological classification, E$=$elliptical, S$=$spiral and I$=$irregular.
}
  \label{table1}
  \tiny
  \centering
    \begin{tabular}{ccccccccccccccccc}
      \hline
     \hline
Target & $z$ & $P_{\rm [CII]}$ &{\sc fwhm}$_{\rm [CII]}$ & $S_{\rm [CII]}$ & $L_{\rm [CII]}$ & ${\rm log_{10}}[L_{\operatorname{IR}}/L_\odot]$ & $T_{\rm dust}$ & $\beta$ & $\alpha_{\operatorname{mid-IR}}$ & \Cii/IR & $r_{\rm eff, 50\%}$ & ${\rm log_{10}(M_\star/M_\odot)}$ & 
${\rm sSFR_{IR}}$ &  12+log$_{10}$(O/H) & BPT & Morph.  \\
 &  & (Jy) & (km\,s$^{-1}$) & (Jy\,km\,s$^{-1}$) & $(\times10^{8} L_\odot)$ &  & $(K)$ &  &  &  ($\times10^{-3}$) & (kpc) &  & (Gyr$^{-1}$) &   & class &  \\
(1) & (2) & (3) & (4) & (5) & (6) & (7) & (7) & (9) & (10) & (11) & (12) & (13) & (14) & (15) & (16) & (17) \\
      \hline
      \hline
G09.v2.102    & 0.073 & 1.90$\pm$0.06      & 356$\pm$22.88 & 849$\pm$46      &  1.70$\pm$0.09 & 11.03$\pm$0.01 & 36$\pm$ 2 & 1.2$\pm$0.2 & 2.10$\pm$0.06 & 1.60 &  3.5 & 10.63$\pm$0.12 & 0.30 & -- & 1 & ES \\ 
G09.v2.103    & 0.041 & 4.45$\pm$0.06      & 220$\pm$0.10 & 1509$\pm$31      &  0.95$\pm$0.02 & 10.28$\pm$0.03 & 25$\pm$ 2 & 1.5$\pm$0.4 & 3.00$\pm$0.18 & 4.96 &  5.0 & 10.48$\pm$0.12 & 0.08 & -- & 1 & S  \\ 
G09.v2.107    & 0.128 & 4.50$\pm$0.09      & 227$\pm$21.31 & 1467$\pm$62      &  9.33$\pm$0.40 & 11.70$\pm$0.01 & 39$\pm$ 1 & 1.4$\pm$0.1 & 2.50$\pm$0.05 & 1.86 &  4.2 & 10.51$\pm$0.13 & 1.84 & 8.923$\pm$0.011 & 0 & I  \\ 
G09.v2.111    & 0.078 & 3.39$\pm$0.07      & 192$\pm$18.32 & 1052$\pm$42      &  2.43$\pm$0.10 & 11.15$\pm$0.01 & 41$\pm$ 2 & 0.9$\pm$0.2 & 2.30$\pm$0.06 & 1.72 &  5.1 & 10.55$\pm$0.13 & 0.48 & 9.147$\pm$0.013 & 0 & ES \\ 
G09.v2.117    & 0.054 & 2.99$\pm$0.06      & 168$\pm$12.24 & 898$\pm$31      &  0.99$\pm$0.04 & 10.42$\pm$0.04 & 22$\pm$ 2 & 1.6$\pm$0.3 & 2.50$\pm$0.18 & 3.75 &  7.6 & 10.70$\pm$0.12 & 0.06 & -- & -- & ES \\ 
G09.v2.137    & 0.044 & 6.05$\pm$0.06      & 284$\pm$8.39 & 2349$\pm$42      &  1.71$\pm$0.03 & 10.54$\pm$0.02 & 34$\pm$ 2 & 1.3$\pm$0.2 & 2.90$\pm$0.17 & 4.90 &  2.5 & 10.13$\pm$0.13 & 0.31 & 9.010$\pm$0.006 & 0 & ES \\ 
G09.v2.167    & 0.078 & 0.46$\pm$0.05      & -- & --      & $<$ 1.25 & 11.05$\pm$0.03 & 50$\pm$ 2 & 0.7$\pm$0.1 & 4.80$\pm$0.20 & $<$1.10 &  2.5 & 10.35$\pm$0.14 & 0.61 & -- & 2 & E  \\ 
G09.v2.170    & 0.051 & 4.52$\pm$0.07      & 195$\pm$12.97 & 1439$\pm$38      &  1.41$\pm$0.04 & 10.33$\pm$0.03 & 21$\pm$ 2 & 2.2$\pm$0.5 & 2.50$\pm$0.13 & 6.54 &  4.4 & 10.26$\pm$0.13 & 0.14 & -- & -- & S  \\ 
G09.v2.175    & 0.070 & 3.03$\pm$0.06      & 312$\pm$17.73 & 1235$\pm$46      &  2.28$\pm$0.09 & 11.18$\pm$0.01 & 42$\pm$ 2 & 1.2$\pm$0.2 & 1.90$\pm$0.05 & 1.51 &  3.4 & 10.48$\pm$0.11 & 0.60 & -- & 1 & E  \\ 
G09.v2.232    & 0.096 & 2.15$\pm$0.07      & 184$\pm$0.01 & 648$\pm$32      &  2.28$\pm$0.11 & 10.93$\pm$0.04 & 23$\pm$ 3 & 2.2$\pm$0.6 & 2.70$\pm$0.20 & 2.67 &  7.5 & -- & -- & -- & 1 & ES \\ 
G09.v2.235    & 0.027 & 3.11$\pm$0.07      & 142$\pm$19.93 & 901$\pm$34      &  0.24$\pm$0.01 & 10.20$\pm$0.01 & 50$\pm$ 4 & 0.3$\pm$0.2 & 1.80$\pm$0.05 & 1.53 &  1.6 &  9.94$\pm$0.11 & 0.22 & 8.665$\pm$0.000 & 0 & ES \\ 
G09.v2.23     & 0.033 & 5.25$\pm$0.05      & 356$\pm$8.34 & 2373$\pm$44      &  0.97$\pm$0.02 & 10.31$\pm$0.02 & 24$\pm$ 2 & 1.6$\pm$0.4 & 2.80$\pm$0.13 & 4.72 &  3.6 & 10.55$\pm$0.11 & 0.07 & 8.935$\pm$0.007 & 0 & ES \\ 
G09.v2.26     & 0.182 & 4.56$\pm$0.13      & 103$\pm$23.42 & 1052$\pm$51      & 13.81$\pm$0.68 & 11.84$\pm$0.02 & 24$\pm$ 1 & 2.1$\pm$0.2 & 2.20$\pm$0.08 & 2.02 &  6.4 & 11.25$\pm$0.12 & 0.46 & 9.142$\pm$0.007 & 0 & EI \\ 
G09.v2.299    & 0.074 & 2.44$\pm$0.06      & 258$\pm$25.34 & 880$\pm$48      &  1.81$\pm$0.10 & 10.98$\pm$0.01 & 45$\pm$ 3 & 0.8$\pm$0.2 & 2.40$\pm$0.08 & 1.91 &  2.3 & 10.54$\pm$0.11 & 0.33 & 9.214$\pm$0.012 & 0 & E  \\ 
G09.v2.38     & 0.059 & 2.94$\pm$0.05      & 527$\pm$16.39 & 1794$\pm$58      &  2.30$\pm$0.08 & 10.70$\pm$0.02 & 22$\pm$ 1 & 2.0$\pm$0.3 & 3.00$\pm$0.17 & 4.55 &  4.6 & 10.89$\pm$0.11 & 0.08 & 9.115$\pm$0.648 & 0 & ES \\ 
G09.v2.42     & 0.055 & 3.89$\pm$0.07      & 215$\pm$17.74 & 1291$\pm$43      &  1.45$\pm$0.05 & 11.09$\pm$0.01 & 38$\pm$ 1 & 1.3$\pm$0.1 & 2.30$\pm$0.04 & 1.17 &  2.8 & 10.46$\pm$0.12 & 0.51 & -- & 1 & E  \\ 
G09.v2.45     & 0.051 & 6.10$\pm$0.06      & 266$\pm$7.28 & 2274$\pm$41      &  2.18$\pm$0.04 & 10.71$\pm$0.01 & 31$\pm$ 1 & 1.3$\pm$0.2 & 3.10$\pm$0.13 & 4.27 &  4.0 & 10.28$\pm$0.13 & 0.32 & 8.901$\pm$0.009 & 0 & S  \\ 
G09.v2.48     & 0.072 & 7.17$\pm$0.07      & 223$\pm$7.67 & 2398$\pm$43      &  4.69$\pm$0.09 & 11.27$\pm$0.01 & 36$\pm$ 1 & 1.4$\pm$0.1 & 2.10$\pm$0.03 & 2.51 &  2.4 & 10.58$\pm$0.13 & 0.59 & -- & 1 & EI \\ 
G09.v2.52     & 0.026 & 4.35$\pm$0.06      & 168$\pm$11.33 & 1331$\pm$33      &  0.33$\pm$0.01 & 10.25$\pm$0.01 & 27$\pm$ 1 & 1.8$\pm$0.3 & 1.90$\pm$0.05 & 1.81 &  2.7 & 10.29$\pm$0.11 & 0.11 & -- & 2 & ES \\ 
G09.v2.55     & 0.054 & 4.28$\pm$0.07      & 296$\pm$15.46 & 1697$\pm$53      &  1.86$\pm$0.06 & 10.57$\pm$0.02 & 23$\pm$ 1 & 2.0$\pm$0.3 & 2.90$\pm$0.11 & 5.05 &  3.3 & 10.81$\pm$0.11 & 0.07 & -- & 1 & ES \\ 
G09.v2.58     & 0.052 & 5.41$\pm$0.06      & 223$\pm$9.54 & 1832$\pm$39      &  1.86$\pm$0.04 & 10.69$\pm$0.02 & 25$\pm$ 2 & 2.0$\pm$0.4 & 2.60$\pm$0.10 & 3.81 &  3.4 & 10.60$\pm$0.12 & 0.15 & 9.086$\pm$0.047 & 0 & ES \\ 
G09.v2.60     & 0.060 & 3.52$\pm$0.05      & 482$\pm$13.53 & 1997$\pm$55      &  2.66$\pm$0.07 & 10.72$\pm$0.02 & 23$\pm$ 2 & 2.0$\pm$0.4 & 2.90$\pm$0.17 & 5.03 &  5.0 & 10.63$\pm$0.12 & 0.15 & -- & 1 & S  \\ 
G09.v2.66     & 0.031 & 6.73$\pm$0.07      & 177$\pm$8.13 & 2090$\pm$35      &  0.73$\pm$0.01 & 10.13$\pm$0.03 & 24$\pm$ 2 & 1.5$\pm$0.3 & 2.60$\pm$0.17 & 5.38 &  3.1 & 10.35$\pm$0.12 & 0.07 & -- & 1 & S  \\ 
G09.v2.76     & 0.107 & 3.61$\pm$0.08      & 230$\pm$18.52 & 1201$\pm$47      &  5.30$\pm$0.21 & 11.22$\pm$0.01 & 36$\pm$ 3 & 0.7$\pm$0.2 & 2.30$\pm$0.07 & 3.21 &  7.9 & 11.07$\pm$0.12 & 0.17 & -- & 2 & EI \\ 
G09.v2.77     & 0.079 & 3.92$\pm$0.06      & 403$\pm$14.68 & 1917$\pm$58      &  4.51$\pm$0.14 & 11.01$\pm$0.03 & 25$\pm$ 2 & 2.0$\pm$0.4 & 2.50$\pm$0.18 & 4.38 &  4.6 & 10.59$\pm$0.13 & 0.32 & -- & 1 & S  \\ 
G09.v2.80     & 0.053 & 3.42$\pm$0.07      & 104$\pm$15.06 & 907$\pm$32      &  0.94$\pm$0.03 & 10.44$\pm$0.02 & 21$\pm$ 2 & 2.1$\pm$0.4 & 2.60$\pm$0.12 & 3.44 &  4.0 & 10.75$\pm$0.12 & 0.06 & -- & 1 & ES \\ 
G09.v2.90     & 0.133 & 1.70$\pm$0.06      & 459$\pm$55.95 & 909$\pm$91      &  6.22$\pm$0.63 & 11.51$\pm$0.02 & 32$\pm$ 3 & 1.6$\pm$0.2 & 2.60$\pm$0.16 & 1.93 &  4.7 & 10.73$\pm$0.11 & 0.73 & -- & 1 & E  \\ 
G09.v2.87     & 0.195 & 1.17$\pm$0.12      & -- & --      & $<$11.38 & 11.92$\pm$0.01 & 39$\pm$ 2 & 1.3$\pm$0.2 & 4.40$\pm$0.98 & $<$1.36 &  3.3 & 10.59$\pm$0.10 & 2.58 & -- & 2 & E  \\ 
      \hline
      \hline
    \end{tabular}
\end{table}
\end{landscape}

\begin{figure}
   \centering
   \includegraphics[scale=0.55]{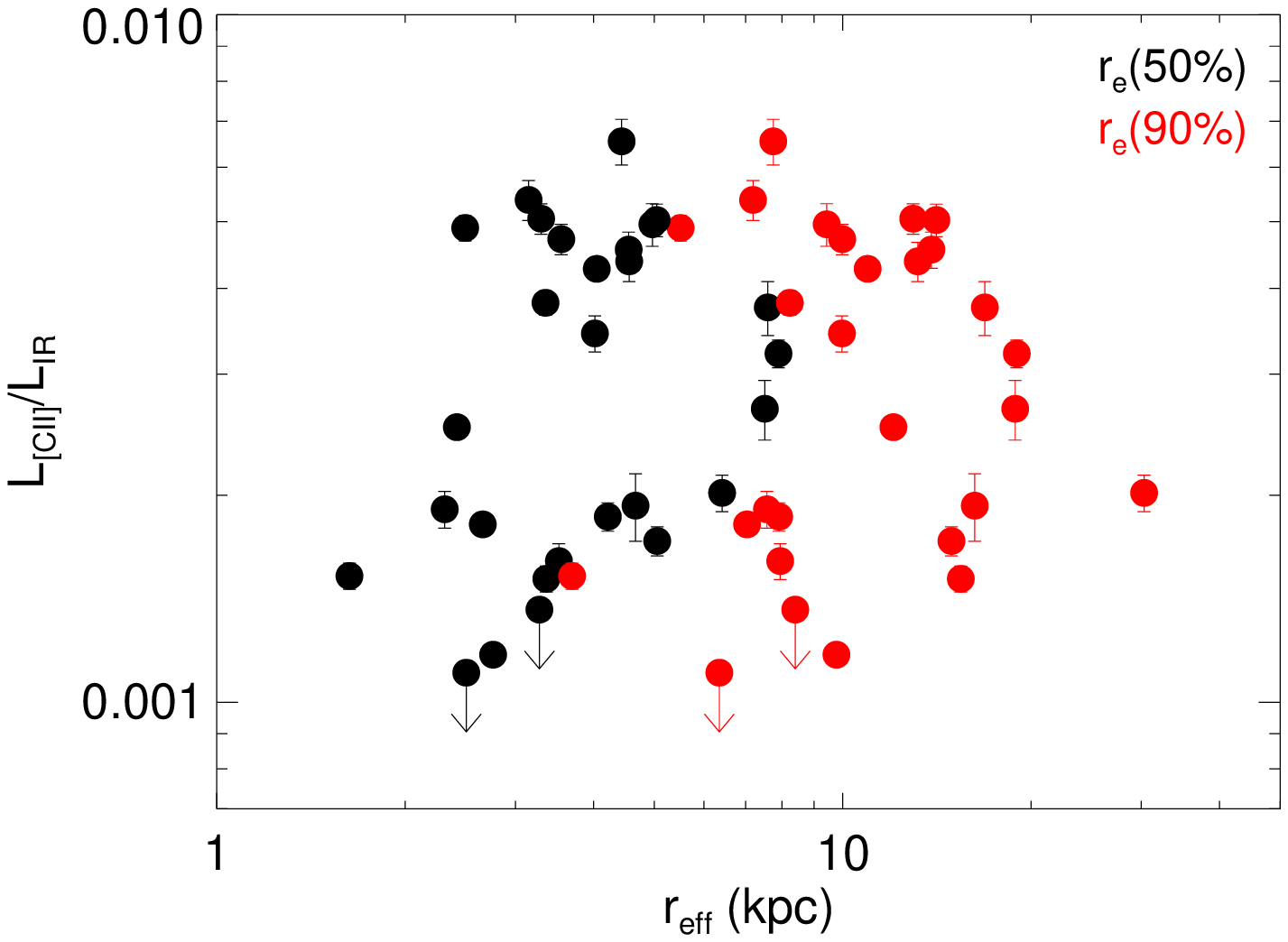}
   \includegraphics[scale=0.55]{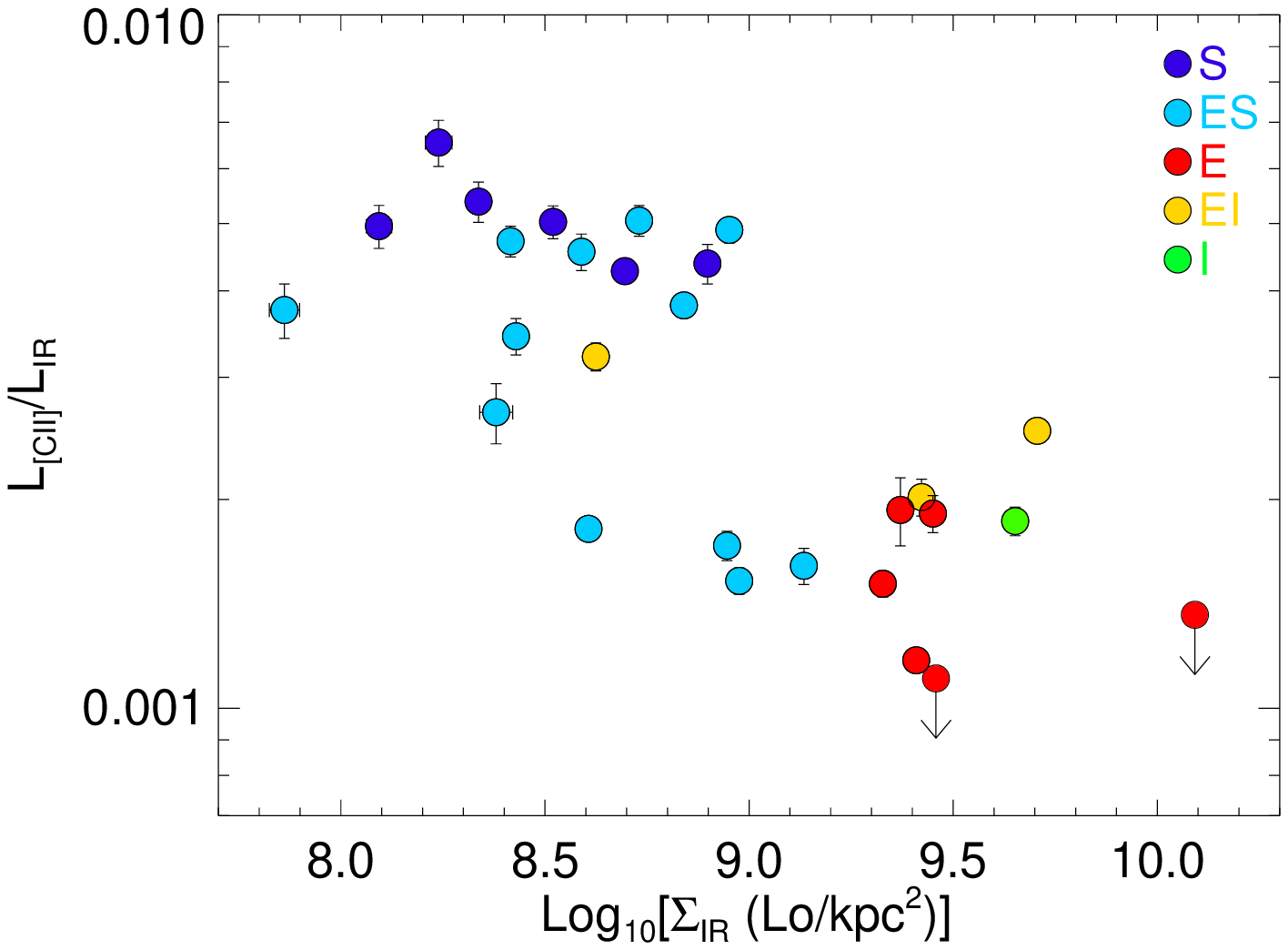}
   \caption{{\it Top:} \Cii/IR luminosity ratio as a function of
     effective radius in SDSS $r$-band (encircled fraction at 50\% in
     black and 90\% in red). {\it Bottom:} The variation of the
     \Cii/IR luminosity ratio as a function of surface brightness
     ($\Sigma_{\rm IR}=L_{\rm IR}/(2\pi r_{\rm eff,50\%}^2)$;
     here $r_{\rm eff,50\%}$ is based on the optical
       $r$-band size and $\Sigma_{\rm IR}$ has not been corrected by
       inclination), colour coded as a function of a simply by-eye
       optical morphological classification; spirals (`S'),
       ellipticals (`E'), irregulars (`I'; including mergers), or
       composite morphologies. This figure clearly shows
     that higher \Cii/IR ratios are preferentially seen
       in galaxies presenting extended disks.}
   \label{image_petro}
\end{figure}

\begin{figure}
   \centering
   \includegraphics[scale=0.55]{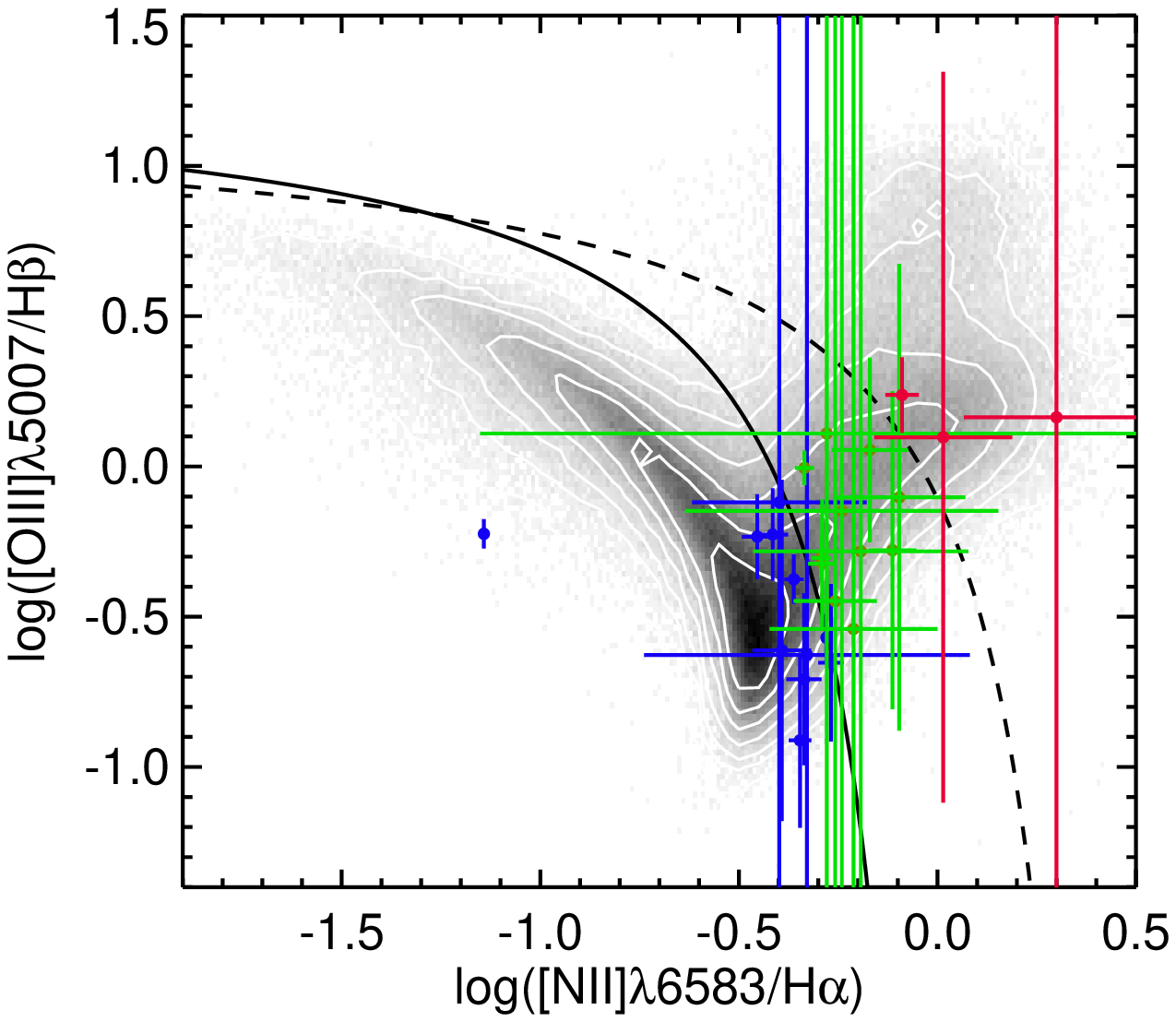}
   \includegraphics[scale=0.52]{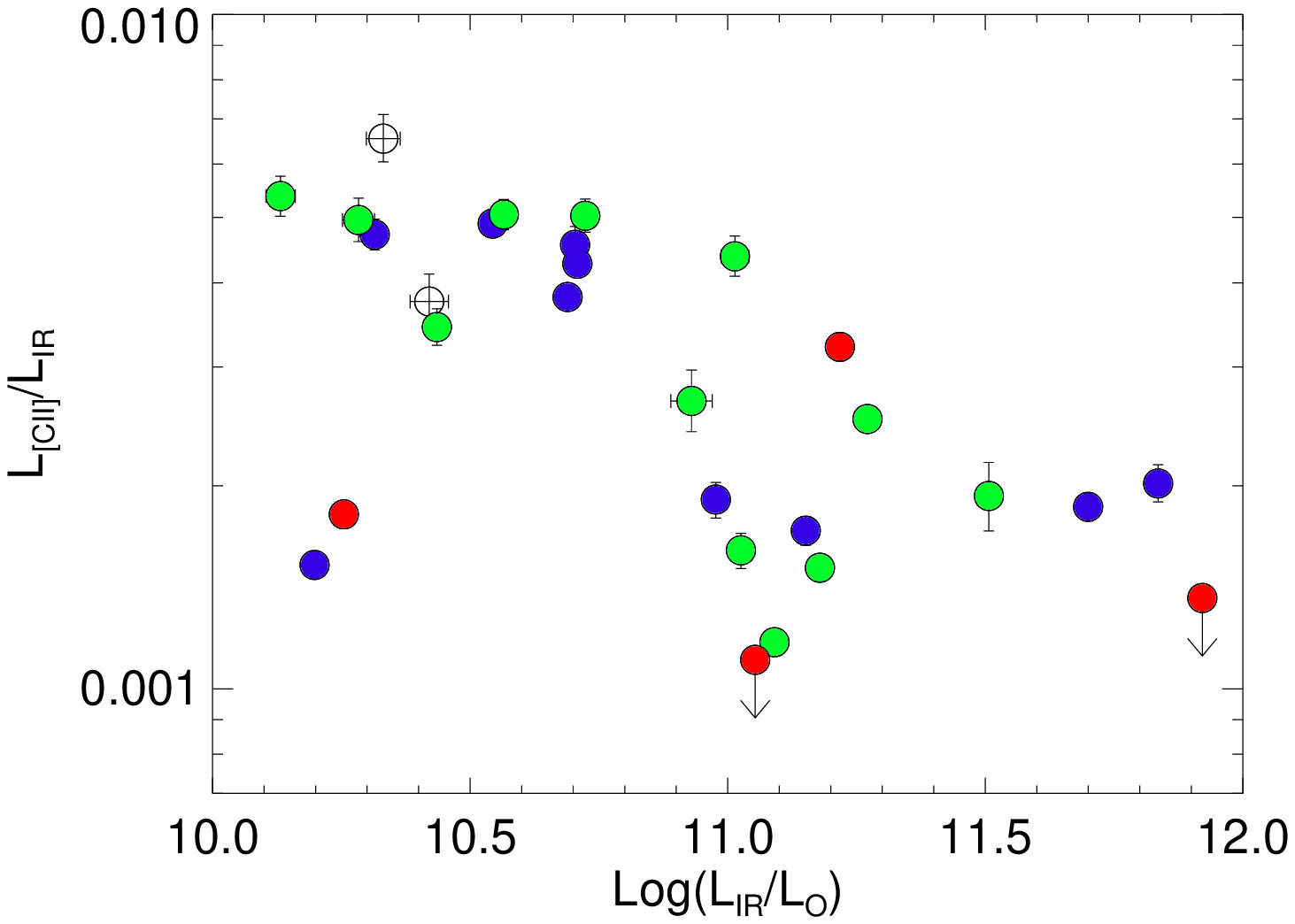}
   \caption{{\it Top:} The nature of our targets as shown by the BPT
     diagram (see \S\,\ref{subsec_optical}). Nine galaxies present
     line ratios with large error bars (0.1\,dex). Galaxies to the
     left of the solid line (taken from \citealt{Kauffmann03}) are
     considered star-forming (blue crosses); those to the right of the
     dashed line (taken from \citealt{Kewley01}) are considered AGNs
     (red crosses); and those between the solid and dashed lines are
     defined as composites (green crosses). The background contours
     and grey scale shows the distribution for the whole spectroscopic
     SDSS sample. {\it Bottom:} The \Cii/IR luminosity ratio as a
     function of IR luminosity, colour-coded by the BPT classification
     (same colours as above). The present data show that in this
     diagram the different populations are indistinguishable from each
     other.}
   \label{bpt_diagram}
\end{figure}

\subsection{Correlations with optical properties}
\label{subsec_optical}

In this section, we explore the optical properties of the
targets. First, we look for morphological features. We downloaded the
corresponding {\sc fits} files $r$-band images from the SDSS
archive\footnote{\sc skyserver.sdss3.org/dr9/en/tools/chart/} (see
Fig.~\ref{fig_postages}). We masked all sources which are not
contributing to the far-IR emission, and created the encircled energy
fraction as a function of a radius centred at the peak source
position. We measured the effective radius at which 50\,\% (and
90\,\%) of the power is encircled, and then transformed this projected
size into a physical scale at the given redshift of the galaxy. These
scales are shown in Fig.~\ref{image_petro} as a function of the
\Cii/IR ratio. No clear correlation is found between these two
observables (Table~\ref{table_correlations}). 

We then performed a
basic visual inspection using the $r$-band SDSS imaging to classify
these galaxies in three populations: ellipticals (`E'), spirals (`S')
and irregulars (`I'). Some galaxies show morphologies which are not
possible to classify in a single population. In these cases we use a
combination of letters, e.g.\ `ES' or `EI', to refer to a prominent
bulge with a disk or an irregular morphology, respectively. Under this
classification, in Fig.~\ref{image_petro} we show that the \Cii/IR
ratio is preferentially higher in galaxies presenting a prominent disk
compared to those which do not present disky morphologies. 

In order to provide an estimate for the IR surface
  brightness of the galaxies, we consider the optical $r$-band radius
  as a proxy for the actual IR radius, so we define $\Sigma_{\rm
    IR}=L_{\rm IR}/(2\pi r_{\rm eff,50\%}^2)$. We reckon this is a
  strong assumption so this value should be interpreted with
  caution. As expected, we find that the previous morphological
  classification is correlated to the surface brightness. Indeed,
those with low $\Sigma_{\rm IR}$ values tend to be those presenting
disk-like morphologies, while those which are classified as pure
elliptical tend to be those with highest brightness in the
sample.

With the aim to identify possible \Cii\ self-absorption
(e.g.\ following \citealt{Gerin15}), we have looked for the properties
of edge-on galaxies within the spiral population (see
Fig.~\ref{fig_postages}). The four identified edge-on galaxies tend to
have a well defined high \Cii/IR ratios of $3.5\,\times\,10^{-3}$, and
do not deviate from the rest of the `spiral-only' population.

To characterise the nature of our galaxies, we have utilised the SDSS
and GAMA spectra to locate them in the ``Baldwin, Phillips \&
Terlevich'' (BPT) diagram (\citealt{Baldwin81}). Emission line
strengths for H$\alpha$, H$\beta$, [N\,{\sc ii}]$\lambda$6583 and
[O\,{\sc iii}]$\lambda$5007 are extracted from spectra using the
GANDALF pipeline (see details in \citealt{Hopkins13}). Only two
galaxies do not have all four lines detected. We show these data in
Fig.~\ref{bpt_diagram}, where our sample is over-plotted on top of the
whole spectroscopic SDSS population. We find that our sample locates
all over the star-forming, AGN and composite (a mix of both)
regions. The main selection criterion, $S_{\rm 160\mu m}>150\,$mJy,
therefore introduces a significant number of low-luminosity AGNs in
the sample. This permits to test if AGN activity plays an important
role in the \Cii/IR ratio. Note that none of our galaxies falls close
to the peak of the distribution of star-forming SDSS galaxies. We show
that under the BPT classification, our sample of star-forming and
composite galaxies do not show significant differences in terms of
\Cii/IR ratios (Fig.~\ref{bpt_diagram}). We note, however, that the
uncertainties in the line flux ratios might be blurring any possible
correlation.

The stellar masses ($M_\star$) for all galaxies are calculated as
described in \citet{Taylor11}, using GAMA catalogue version v08. The
stellar mass estimates were derived by fitting their SEDs
(\citealt{Bruzual03}) to the SDSS $ugriz$ imaging -- data which have
been re-processed by the GAMA team (\citealt{Hill11}). The dust
obscuration law applied was that of \citet{Calzetti00}, and a
\citet{Chabrier03} IMF was assumed. The stellar masses
  are determined by integrals that are weighted to the probability of
  each SED fit. This has been performed to all galaxies regardless of
  the nature (star-forming/AGN) defined by the BPT diagram. We have
  not applied any conversion factor to convert from Chabrier to Kroupa
  IMF because the variation is negligible compared to the measured
  errors. \citeauthor{Taylor11} demonstrate that the relation between
($g$--$i$) and $M_\star/L$ offers a simple indicator of the stellar
masses. To account for aperture effects, a correction based on the
S\'ersic fit to the surface brightness profiles is applied to the
stellar masses (see \citealt{Taylor11,Kelvin12}).

\begin{figure}
   \centering
   \includegraphics[scale=0.55]{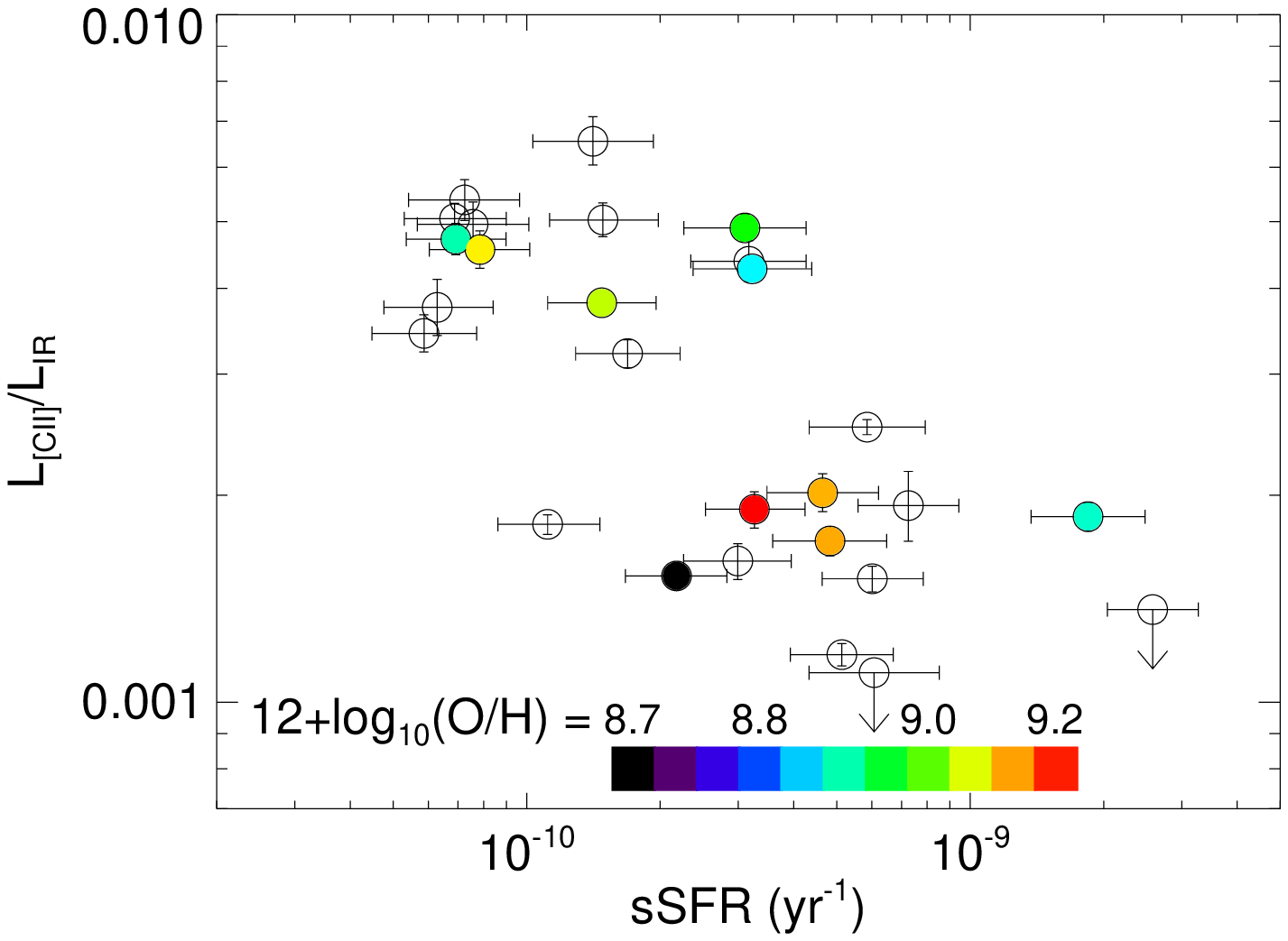}
   \includegraphics[scale=0.55]{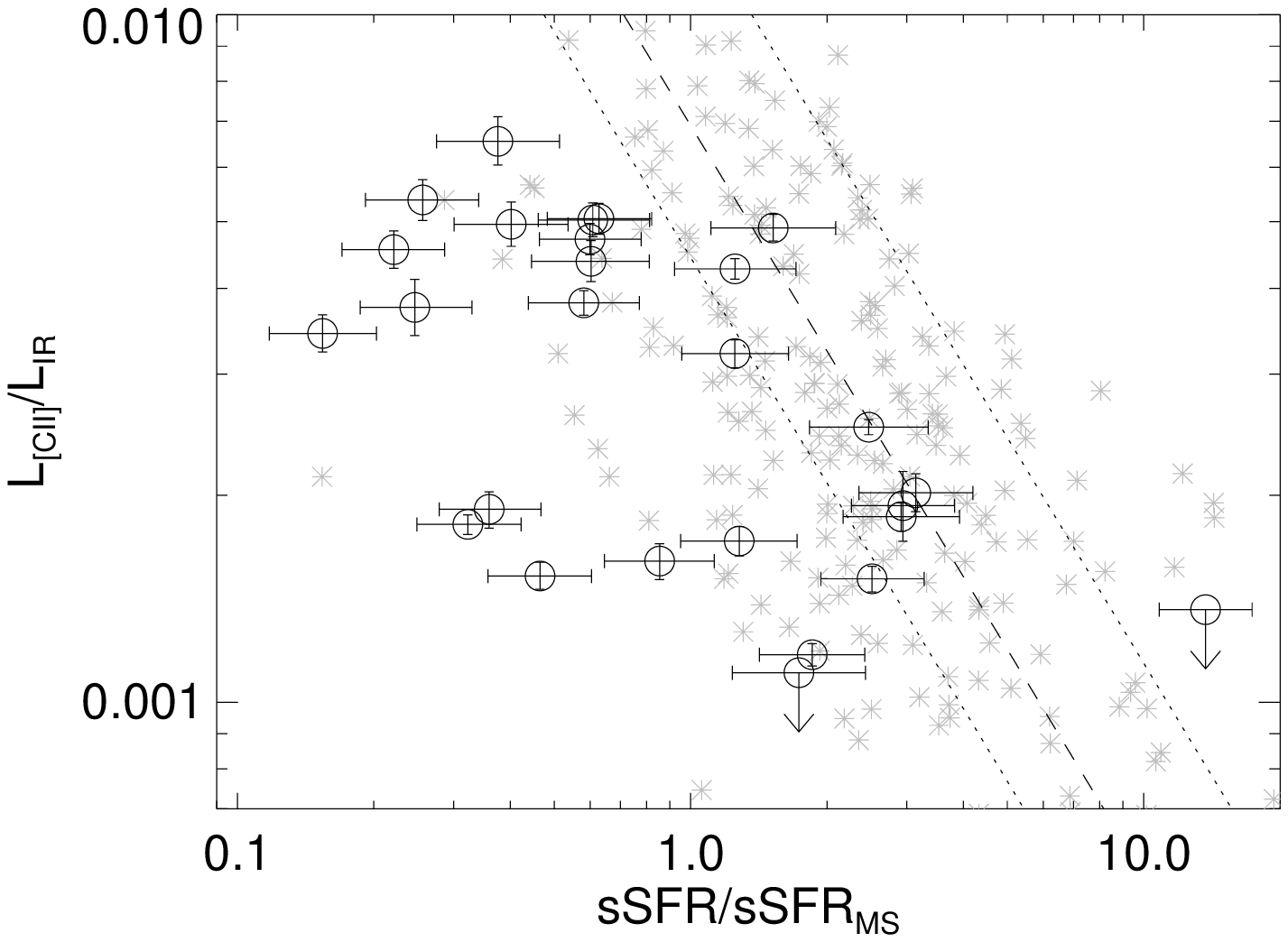}
   \caption{{\it Top:} The \Cii/IR luminosity ratio as a function of
     the sSFR, derived using the observed IR luminosity and stellar
     masses from SEDs fits to SDSS $u$$g$$r$$i$$z$ photometry (see
     \S\,\ref{subsec_optical} for details). Colour-coded we show
     metallicity measurements (using \citealt{Tremonti04}'s relation)
     for only those sources which are identified as star-forming in
     the BPT diagram (Fig.~\ref{bpt_diagram}). {\it Bottom:} The same
     figure but dividing the sSFR values by the redshift dependent
     `main-sequence' defined by \citet{Elbaz11}. In
       light grey asterisks we show the galaxy sample presented by
       \citet{Diaz-Santos13}. The dashed line shows the correlation
       defined by \citeauthor{Diaz-Santos13} (see their Eqn.~5 but
       converting $L_{\rm IR}=1.75\times L_{\rm FIR}$), including a
       $\pm$\,0.26\,dex range of dispersion (dotted lines). Note how
     our sample is more representative of `normal' galaxies (with
     respect to the `main-sequence') compared to that presented by
     \citet{Diaz-Santos13}.}
   \label{sSFR_vs_deficit}
\end{figure}

We use the SFR derived from the bolometric IR
  emission, following \citet{Kennicutt98}. In combination with the
  stellar mass estimate we obtain the specific-SFRs (sSFR\,$={\rm
    SFR_{\rm IR}}/M_\star$). We find the sSFR seems to anti-correlate
  with the \Cii/IR luminosity ratios (see Fig.~\ref{sSFR_vs_deficit}),
  hence those galaxies presenting lower \Cii/IR ratios are
  preferentially passing through more violent bursts of
  star-formation. In order to remove a possible dependency on redshift
  and follwing \citet{Diaz-Santos13}'s work, we have divided our sSFR
  estimates by the one expected at the `main-sequence' defined by
  \citet{Elbaz11} (at the given redshift). We find that our sources
  are centred at the main sequence value but cover a wide range of two
  orders of magnitude around it (see Fig.~\ref{sSFR_vs_deficit}). If
  we compare our sample to \citeauthor{Diaz-Santos13}'s, our sample
  is more representative of `normal' star-forming galaxies than
  theirs.

For those galaxies of which are defined as star-forming in the BPT
diagram (Fig.~\ref{bpt_diagram}), we measure reliable metallicities
using the {\sc o3n2} index of \citet{Pettini04}, and converting those
values to the calibration of \citet{Tremonti04}, following
\citet{LaraLopez13}. We find that these star-forming galaxies present
metallicities between $12+{\rm log_{10}(O/H)} = 8.7-9.2$. For
comparison, Solar metallicity is 8.91. In low metallicity environments
($12+{\rm log_{10}(O/H)} < 8.1$), the IR emission is expected to drop
but the \Cii\ emission would remain almost invariant, therefore the
\Cii/IR ratio is expected to be higher in low metallicity environments
(see also \citealt{Rubin09, deLooze14}). Nevertheless, our sample does
not probe these lower metallicities (mainly due to dust selection
criterion) and so we cannot investigate correlation of \Cii\ with
metallicities.

\section{Discussion}

For many years it has been a great challenge to explain why there is a
decrease in the \Cii/IR luminosity ratio towards bright ($L_{\rm
  IR}>10^{11}L_\odot$) IR luminosities. It is well established that
this ratio has an intimate dependency on the strength of the radiation
field ($<$$G_{\rm O}$$>$) and the density ($n_e$) of the ISM. As
explained in \S\,\ref{sec_intro}, the \Cii\ emission comes from a
whole range of different ISM states; ionised, atomic and/or
molecular. Unfortunately, given the low resolution of our {\it
  Herschel} observations, these different components are impossible to
disentangle directly from imaging. Fibre-based optical spectroscopy
cannot spatially separate these components neither. 

First of all, we look for possible biases introduced by the selection
criteria used to construct our galaxy sample. We identify that one of
the most important criterion is the flux density $S_{\rm 160\mu
  m}>150\,$mJy threshold, which introduces a strong selection effect
on the redshift-luminosity distribution of the targets (see
Fig.~\ref{fig_threshold}). Our targets span a redshift range between
$z=0.02$ and $0.2$, so approximately 2.2\,Gyr of cosmic
time. Based on the measured sSFR, the galaxies in our
  sample double their stellar masses in scales of ${\rm
    sSFR}^{-1}\approx0.3-20\,$Gyr (median 4\,Gyr; see
  Fig.~\ref{sSFR_vs_deficit}), values which are in most cases longer
  than 2.2\,Gyr. In this work, we consider that for most of our
  targets the evolution is short in comparison to the
    redshift slide, i.e.\ a galaxy at fixed luminosity will be
  behaving the same at $z=0.02$ as at $z=0.2$. Only a fourth of the
  targets double their stellar masses in scales which are shorter than
  2.2\,Gyr (those ongoing more violent star-bursts). This assumption
is necessary to alleviate the luminosity-redshift dependency seen in
Fig.~\ref{fig_threshold}. In terms of the ISM
  evolution using large samples of galaxies, \citet{Lara-Lopez09}
finds no variation of the metallicity properties in this redshift
range, nevertheless the dust mass density evolves strongly as a
function of redshift, incrementing by a factor of two from $z=0.02$ to
$0.2$ (\citealt{Dunne11}). For the purposes of this work, we assume
the flux density threshold at $S_{\rm 160\mu m}>150\,$mJy does not
introduce biases on the results.

Significant differences are found between galaxies presenting high and
low \Cii/IR luminosity ratios (see Table~\ref{table_correlations}) --
understood in this study as greater or lower than
$\sim$\,$2.5\,\times\,10^{-3}$. We find that galaxies
with high ratios:
\begin{itemize} 
\item have cold dust temperatures, preferentially lower than 30\,K
  (see Fig.~\ref{fig_Tdust}). This evidence indicates that these
  galaxies present a prominent IR component coming from extended
  ISM regions rather than compact ones located at the vicinity of
  powerful star-forming nuclear regions;
\item have a high {\it WISE} flux density ratio of $0.5\lesssim S_{\rm
  12\mu m}/S_{\rm 22\mu m}\lesssim1.0$ (see
  Fig.~\ref{deficit_WISE}). This range of mid-IR ratios tend to be
  associated to normal {\it Hubble}-type spiral galaxies, a ratio
  induced by a combination of spectral features within the broad {\it
    WISE}\, $12\,\mu$m and 22\,$\mu$m filters: prominent PAH emission
  lines (not suppressed by strong radiation fields), a weak 10\,$\mu$m
  Silicate absorption band (indicating moderate extinction levels),
  and a $\sim$\,22\,$\mu$m spectra which does not seem to be dominated
  by powerful hot dust emission, e.g.\ a young and violent star-burst or
  an AGN torus;
\item have preferentially lower surface brightness ($\Sigma_{\rm
  IR}\lesssim10^9\,L_\odot/{\rm kpc^2}$) as shown in
  Fig.~\ref{image_petro} (see significance in
  Table~\ref{table_correlations}). We stress this
    estimate uses the effective $r$-band radius as a proxy for the IR
    extension. If the mean free path of the far-UV photons is large,
  then the strength of the radiation fields would be directly
  proportional to $\Sigma_{\rm IR}$ (\citealt{Wolfire90}). If this is
  the case, this suggests that galaxies with weaker radiation fields
  produce higher \Cii/IR luminosity ratios. Nevertheless, given the
  lack of correlation between \Cii/IR and $r_{\rm eff}$ (see
  Table~\ref{table_correlations}), this behaviour could be a
  manifestation of the correlation with $L_{\rm IR}$ instead;
\item have preferentially disk-like morphologies. Those galaxies which
  are classified as spirals (`S'), without prominent bulges, are those
  with the highest \Cii/IR ratios ($\,\sim\,4\times10^{-3}$; see
  Fig.~\ref{image_petro}), similar to that found by
  \citet{Diaz-Santos13} while looking at the pure star-forming LIRGs
  in their sample. These evidences suggest that galaxies with high
  \Cii/IR ratios evolve quiescently rather than triggered by a major
  merger event with a subsequent powerful nuclear star-burst;
\item present a wide range of
  sSFR\,$\approx5\times10^{-11}-3\times10^{-9}$\,yr$^{-1}$ (IR-based),
  hence the mechanism controlling the \Cii\ emission does not seem to
  relate to the efficiency of converting gas into stars. At constant
  SFR, these galaxies range from $3.3$ to $20$\,Gyr to double their
  stellar masses. These results do not agree with the \Cii/IR versus
  sSFR/sSFR$_{\rm MS}$ correlation (see Fig.~\ref{sSFR_vs_deficit})
  found by \citet{Diaz-Santos13}, probably because we observe more
  `normal' galaxies (relative to the `main-sequence' defined by
  \citealt{Elbaz11}) than their local LIRGs sample,
  and also given by the fact that our sample does not
    include a significant number of galaxies with low \Cii/IR ratios
    that permit to evaluate the correlation at higher sSFR
    levels. Our data might suggest an evidence for a plateau in
  \citeauthor{Diaz-Santos13}'s correlation at sSFR/sSFR$_{\rm MS}<1$,
  probably induced by a different and more inefficient star-formation
  mechanism controlling the \Cii/IR ratio
  (\citealt{Daddi10,Gracia-Carpio11}).
\end{itemize} 

\subsection{The strength of the radiation field}

Modelling the \Cii\ emission as coming from PDRs, we suggest that one
of the main parameters responsible in controlling the \Cii/IR ratio is
the strength of the far-UV radiation field ($<$$G_{\rm O}$$>$). This
is supported by the significant correlation found between dust
temperatures, and $\Sigma_{\rm IR}$, with the $L_{\rm [CII]}/L_{\rm
  IR}$ luminosity ratio. Higher dust temperatures suggest higher
radiation fields generated by higher SFR surface densities, which
might create large ionised complexes, especially expected in those
galaxies with high IR luminosities $11<{\rm log_{10}}(L_{\rm
  IR}/L_\odot)<12$. These more extreme conditions could easily change
the dominant ISM state responsible for the bulk of the \Cii\ emission
(e.g.\ \citealt{Diaz-Santos13}).

Actually, the far-UV radiation field (produced by O and B stars) is
one of the main contributors to the heating of the gas via the
photo-electric effect on dust grains.
In the case of soft radiation fields, the ejection rate of
photo-electrons from dust decreases (e.g.\ \citealt{Spaans94}), while
in the case of strong radiation fields, the dust grains become
positively charged, increasing the potential well that the
photo-electrons need to overcome, and thus reducing the input energy
transferred to the gas by photo-electrons (\citealt{Tielens85,
  Malhotra97, Malhotra01, Luhman03}).

We suggest that the \Cii/IR ratio is controlled by the strength of the
far-UV radiation fields, hence the decrement of the \Cii\ line
with respect to IR emission is most probably due to an increment of
positively charged dust grains (higher dust temperatures), which
reduces the efficiency of the far-UV radiation field in transferring
energy into the gas.

\citet{Negishi01} found that the $<$$G_{\rm O}$$>$/$n_e$ (where $n_e$
is the density of electrons of the ISM) ratio does not drive the
\Cii/IR ratio but they suggest that high gas densities play an
important role in controlling the \Cii\ emission. Morphologically
speaking, our analysis shows that galaxies with low \Cii/IR ratios
tend to have prominent bulges in nuclear regions, i.e.\ probably
suggesting that gas density plays an important role in the \Cii/IR
ratio.  However, we were unable to identify a clear correlation
between \Cii/IR ratio and sSFR, indicating that more efficient SFR in
compact regions is probably not controlling the
  \Cii/IR ratio (at least at the parameter space explored by this
  work). Unfortunately, with the available data presented in this
work, we are unable to separate the intimate relation between the
strength of the radiation field and the density of the ISM. To
separate both parameters we require \Cii\ together with another
emission line, such as the fine transitions of [N\,{\sc ii}], [O\,{\sc
    i}] and [C\,{\sc i}] or rotational transitions of CO to properly
determine the physical conditions of the ISM
(e.g.\ \citealt{Wolfire89, Hailey-Dunsheath10}).

\subsection{The ISM origin of the \Cii\ emission}

We show that galaxies presenting high \Cii/IR ratios have
relatively cold dust temperatures, have dominant disk-like
morphologies, and low surface brightness, evidences that indicate
relatively weaker far-UV radiation fields. As previously shown by
\citet{Pineda14}, the origin of the \Cii\ emission in the Milky Way (a
`normal' spiral galaxy) is not only from cold PDRs, but includes also
contributions of the same order from ionised gas, diffuse atomic gas
and CO-dark H$_2$. It is expected that weaker radiation fields would
imply smaller complexes of ionised gas. We argue that sources with
high \Cii/IR ratios might not only emit their \Cii\ luminosity from cold PDRs
but also from the diffuse and extended atomic ISM phase. Without a
H$_2$ tracer (e.g.\ CO lines), we are unable to prove this statement,
although it points out to the difficulties in understanding the origin
of the \Cii\ emission with a single far-IR line detection.

\subsection{Old stellar populations contributing to the cold IR emission}

The \Cii\ and IR luminosities are intimately related to the
star-formation process. In this section we explore if the
`\Cii-deficit' could be a manifestation of an inclusion of an IR
emitting component which is not related to the star-formation, but to
old stellar populations. Actually, the IR SED component coming from
dust heated by old stellar populations is cold, diffuse and is
predominantly emitted at long $>200\,\mu$m wavelengths. We have
explored the possibility that the 500\,$\mu$m luminosity could
correlate with the \Cii/IR ratio, although no clear trends are
observed (see Table~\ref{table_correlations}). If a prominent
cirrus-like emission is present in these galaxies, the Rayleigh-Jeans
would tend to have flatter spectra (between 250 and 500\,$\mu$m),
hence would tend to bias the fitted dust emissivity index (see also
Appendix~\ref{AppendixA}) -- this parameter basically controls the
slope in the Rayleigh-Jeans regime. With the available data it is not
possible to distinguish different physical dust properties from a
strong cirrus component. Nevertheless, it is worth mentioning that 
the anti-correlation found between best fit $\beta$ and the \Cii/IR
luminosity ratio (Fig.~\ref{fig_Tdust}) suggests that it is
unlikely that the bolometric IR luminosity is dominated by a cold
cirrus component, hence responsible for the decrement seen in \Cii/IR
as a function of IR luminosity. This is in agreement with the
selection criterion $S_{\rm 160\mu m}>150\,$mJy which prefers galaxies
dominated by star-forming heating.

\subsection{Self absorption or optically thick}

Extraordinarily large column densities are required to make the
\Cii\ emission optically thick. Self absorption has been employed to
explain the \Cii\ emission from AGN-dominated systems, like Mrk\,231
(\citealt{Fischer10}) where H\,{\sc i} column densities could be
higher than 10$^{22}\,$cm$^{-2}$. For most star-burst galaxies this
effect is expected to be small, especially at the luminosity range
explored by this work $10<{\rm log_{10}} (L_{\rm IR}/L_\odot)
<12$. On the other hand, \citet{Gerin15} have recently
  shown that on the plane of the Milky Way the presence of foreground
  absorption may completely cancel the emission from a background
  far-IR emitter in medium spectral resolution data, suggesting that
  spectra should be taken at high spectral resolution, e.g.\ using
  HIFI (\citealt{deGraauw10}) rather than PACS, to interpret correctly the
  \Cii\ emission, therefore the \Cii/IR ratio.

In order to explore this idea, we have identified all four edge-on
galaxies in our sample as possible optically thick \Cii\ candidates
(along the line of sight). We find that all four galaxies are at the
high end of the \Cii/IR luminosity ratio distribution. This result
suggests that the \Cii\ emission might not be self-absorbed, at least
by the disk, where most of the H\,{\sc ii} regions and surrounding
PDRs are placed. 

Supposing that the whole \Cii\ emission of the disk (mostly
PDR-related) is absorbed, then the observed high \Cii/IR ratios should
come from a diffuse \Cii\ component preferentially located above/below
the disk. Note also that these edge-on galaxies tend to show line
FWHMs which are $\gtrsim\,$200\,km\,s$^{-1}$ (see Table~\ref{table1}),
helping the line emission to escape from the disk.

\subsection{IR emission contaminated by an AGN}

It has been previously shown that sources harbouring an AGN have lower
\Cii\ to IR luminosity ratios (e.g.\ \citealt{Sargsyan12}). In
Fig.~\ref{bpt_diagram}, we have classified star-forming galaxies from
AGNs using the BPT diagram. Error bars are large, although we find that
star-forming, composite and AGN populations are indistinguishable in
terms of \Cii/IR luminosity ratios, possibly suggesting that the
presence of an AGN might be playing a local but not a global role on
the \Cii/IR luminosity ratios -- at least at the low AGN luminosities
presented in this work. This is supported by \citet{Diaz-Santos13} as
they found that AGNs selected by a simple PAH equivalent width
threshold do not modify significantly the \Cii/IR luminosity ratio.

\section{Conclusion}

We have used recent PACS spectroscopic \Cii\ observations to describe
its relation to the IR luminosity in a sample of 28 galaxies selected
from the $H$-ATLAS survey. This sample has high-quality IR
photometry from {\it WISE}, {\it IRAS} and {\it Herschel}, with the
addition of unambiguous photometry and spectroscopy from the SDSS and
GAMA surveys.

In summary, after an exploration over a wide multi-wavelength
parameter space, we have identified the following correlations. We
find that galaxies with high $L_{\rm [CII]}/L_{\rm
  IR}>2.5\times10^{-3}$ luminosity ratios tend to: have $L_{\rm
  IR}<10^{11}\,L_\odot$, dust temperatures lower than 30\,K, high {\it
  WISE} colours in the range $0.5<S_{\rm 12\mu m}/S_{\rm 22\mu
  m}<1.0$, present disk-like morphologies, have low surface brightness
$\Sigma_{\rm IR}\approx10^{8-9}\,L_\odot/{\rm kpc^2}$
(using the $r$-band effective radius), and got a range
  of star-formation rate efficiencies
  (sSFR\,$\approx$\,$0.05-3$\,Gyr$^{-1}$).

Assuming that the physical properties of star-forming galaxies, at
fixed luminosity, are the same at $z=0.02$ and $0.2$ (the range of
redshift of our galaxy sample), and based on the correlations found
between the \Cii/IR luminosity ratio and the dust temperature (and
$\Sigma_{\rm IR}$), we conclude that the most probable parameter
controlling the \Cii/IR luminosity ratio is the strength of the
radiation field (averaged over the entire galaxy) -- probably
inducing an increment of the positive charge of dust grains that has
an effect in the effective energy deposited by the radiation field in
the electrons extracted from dust grains.

The lack of correlation between galaxies with high \Cii/IR luminosity
ratios ($>3\times10^{-3}$) and sSFR values suggests that the
efficiency to convert gas into stars (e.g.\ in high density
environments) is not playing a dominant role in the line to continuum
behaviour, contrary to the correlation found by \citet{Diaz-Santos13}.
We are probably observing a plateau for the correlation at lower
SFR/SFR$_{\rm MS}<1$ ratios, maybe product of a different star-forming
mechanism that controls the \Cii/IR ratio.

We find that the \Cii\ deficit is unlikely to be a manifestation of
optically thick \Cii\ emission, as evidenced by the high \Cii/IR
luminosity ratios found in edge-on spiral galaxies. On the other hand,
the analysis we performed to characterise the nature of our galaxies,
using the BPT diagram, is not conclusive as all star-forming galaxies,
AGNs and composite populations do not clearly distinguish from each
other in terms of the \Cii/IR luminosity ratio. We conclude that at
least at the AGN luminosities shown by our sample, AGN activity does
not seem to play a dominant role in the \Cii\ deficit.

\section{Acknowledgments}

E.\ Ibar acknowledges funding from CONICYT FONDECYT postdoctoral
project N$^\circ$:3130504. RJI, LD and SM acknowledge support from the
European Research Council in the form of Advanced Investigator
programme, COSMICISM. CF acknowledges funding from CAPES
(proc.\ 12203-1). We would like to thank the anonymous referee for the
helpful comments. The {\it Herschel}-ATLAS is a project with {\it
  Herschel}, which is an ESA space observatory with science
instruments provided by European-led Principal Investigator consortia
and with important participation from NASA. The H-ATLAS website is
http://www.h-atlas.org/. PACS has been developed by a consortium of
institutes led by MPE (Germany) and including UVIE (Austria); KU
Leuven, CSL, IMEC (Belgium); CEA, LAM (France); MPIA (Germany);
INAF-IFSI/OAA/OAP/OAT, LENS, SISSA (Italy); IAC (Spain). This
development has been supported by the funding agencies BMVIT
(Austria), ESA-PRODEX (Belgium), CEA/CNES (France), DLR (Germany),
ASI/INAF (Italy), and CICYT/MCYT (Spain). GAMA is a joint
European-Australasian project based around a spectroscopic campaign
using the Anglo-Australian Telescope. The GAMA input catalogue is
based on data taken from the Sloan Digital Sky Survey and the UKIRT
Infrared Deep Sky Survey. Complementary imaging of the GAMA regions is
being obtained by a number of independent survey programs including
GALEX MIS, VST KiDS, VISTA VIKING, {\it WISE}, {\it Herschel}-ATLAS,
GMRT and ASKAP providing UV to radio coverage. GAMA is funded by the
STFC (UK), the ARC (Australia), the AAO, and the participating
institutions. The GAMA website is http://www.gama-survey.org/. This
work has been developed thanks to TOPCAT software
(\citealt{Taylor05}).

\bibliographystyle{mn2e} \bibliography{CII_paper_v8_ArXiV}

\appendix
\section{\\Testing fitted parameters} \label{AppendixA}

As described in \S\,\ref{results_section}, our SED fitting approach
includes a non-standard method. The inclusion of the power-law in the
mid-IR forces the slope of the modified black-body emission at
$\sim$\,100--200\,$\mu$m, possibly introducing a bias on the derived
dust temperature or dust emissivity index. For this reason we repeated
the SED fitting approach excluding the mid-IR slope, leaving just the
modified black body (MBB from Eqn.\ref{mbb_eqn}) component. For these
purposes we just use the {\it Herschel} photometry, i.e.\ the 100,
160, 250, 350 and 500\,$\mu$m data points. We also restricted, between
1.5 and 2.5, the range of possible values for the dust emissivity
index.

\begin{figure}
   \centering
   \includegraphics[scale=0.55]{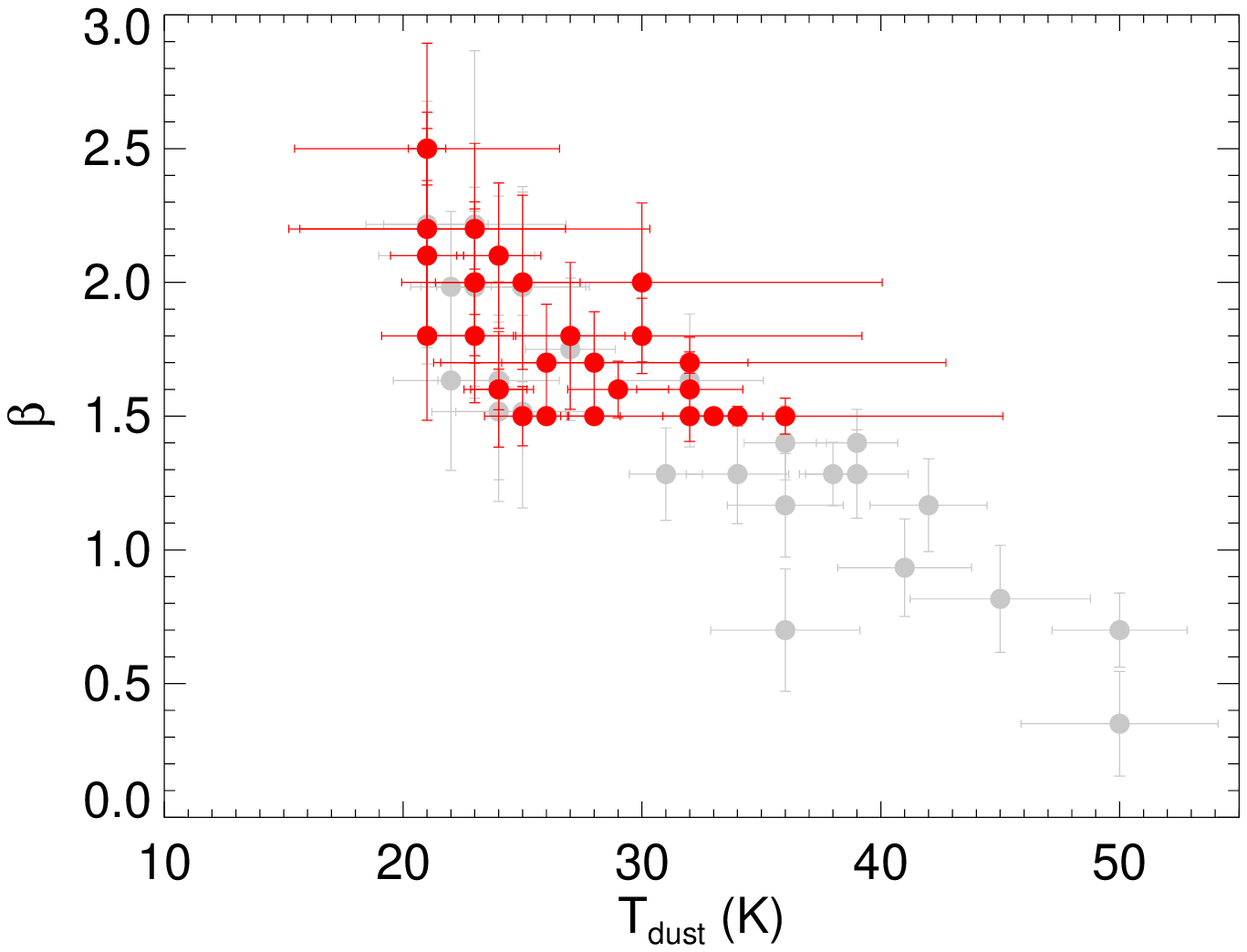}
   \includegraphics[scale=0.55]{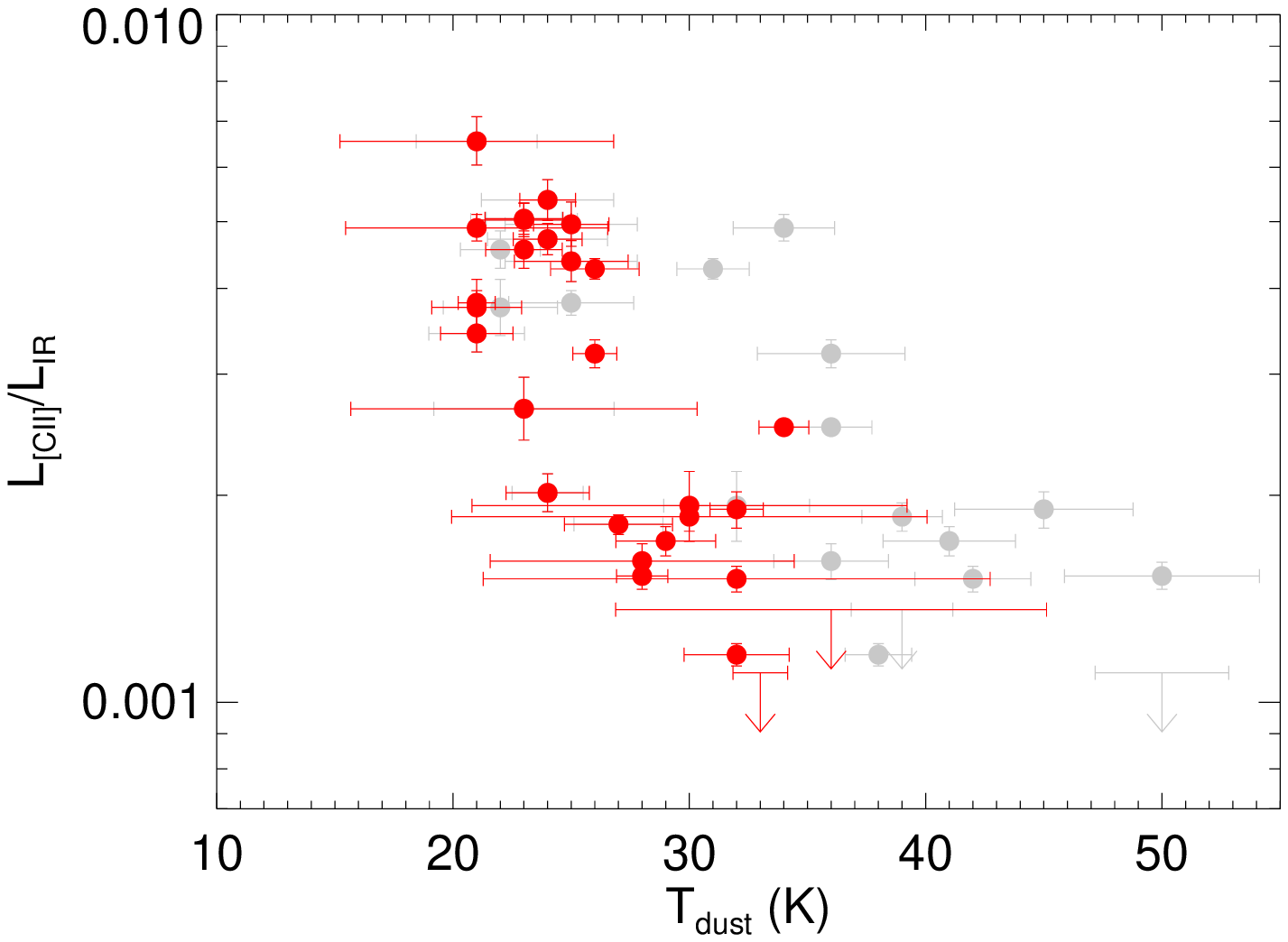}
   \caption{{\it Top:} A comparison between derived parameters,
     $T_{\rm dust}$ and $\beta$, using two different SED fitting
     approaches. In grey, we show the values obtained following the
     method described in \S\,\ref{results_section}. In red, we present
     the values obtained by fitting a modified black body emission
     using only the {\it Herschel} 100--500\,$\mu$m photometry (as
     usually performed in previous {\it Herschel}-based studies). {\it
       Bottom:} The \Cii/IR luminosity ratio as a function of fitted
     dust temperature. Colours are the same as in the top figure. This
     comparison shows that the dependency for \Cii/IR as a function of
     $T_{\rm dust}$ (seen in Fig.\,\ref{fig_Tdust}) remains with this
     different SED-fitting approach.}
   \label{image_comparison_ft}
\end{figure}

The first thing to note is that these new fits are unable to describe
the high-frequency part of the spectra, hence we cannot use them to
get bolometric IR measurements. These new fits, however, show a clear
difference between derived parameters. On average, we find that with
this new SED-fitting method, the $T_{\rm dust}$ decreased by
$\sim$\,6\,K while $\beta$ increased by $\sim$\,0.5 -- parameters
which are well known to be correlated (e.g.\ \citealt{Shetty09,
  Smith13}). These results are shown in
Fig.~\ref{image_comparison_ft}.  This clearly demonstrate that
converting from fitted parameters to `physical' parameters should be
taken with great caution.

Even though a significant difference is seen between derived
parameters coming from these two different SED fitting approaches, the
previous trend seen in \Cii/IR luminosity ratio as a function of dust
temperature remains (see Fig.~\ref{image_comparison_ft}). This
confirms, again, that high \Cii/IR ratios are associated to galaxies
dominated by cold dust emission.

\label{lastpage}

\end{document}